\documentclass[structabstract]{aa}  
\usepackage{multirow,bigdelim}
\usepackage{graphicx}
\usepackage[round]{natbib}
\usepackage{url}
\usepackage{amsmath}
\usepackage[ table ]{ xcolor }
\usepackage{ctable}
\bibpunct{(}{)}{;}{a}{}{,} % to follow the A&A style
\usepackage{txfonts}
\begin{document}
   \title{A high-resolution mm and cm  study of the obscured LIRG NGC~4418}

   \subtitle{A compact obscured nucleus fed by in-falling gas?}

   \author{F. Costagliola
          \inst{1,2}
	\and
          S. Aalto\inst{2}
	\and
	  K.~Sakamoto\inst{3}
	\and
	  S.~Mart{\'i}n\inst{4}	
	\and
	  R.~Beswick\inst{5}
	\and
          S.~Muller\inst{2}
 	\and
	H.-R.~Kl\"ockner\inst{6,7}
          }

   \institute{Istituto de Astrof{\'i}sica de Andaluc{\'i}a, Glorieta de la Astronom{\'i}a, s/n, 18008, Granada, Spain,
        \email{costagli@iaa.es}
	\and Chalmers University of Technology, Onsala Space Observatory, SE-439 92 Onsala, Sweden
  \and Academia Sinica, Institute of Astronomy and Astrophysics, P.O. Box 23-141, Taipei 10617, Taiwan
         \and European Southern Observarory, Alonso de C{\'o}rdova 3107, Vitacura, Casilla 19001, Santiago 19, Chile
	 \and Jodrell Bank Centre for Astrophysics,  Alan Turing Building, School of Physics and Astronomy, The University of Manchester, Manchester M13 9PL, UK
	 \and University of Oxford, Denys Wilkinson Building, Oxford OX1 3RH, UK
	\and Max-Planck-Institut für Radioastronomie, Auf dem Hüegel 69, 53121 Bonn, Germany
             }

   \date{}

% \abstract{}{}{}{}{} 
% 5 {} token are mandatory
 
  \abstract
   {Understanding the nature of the power source in luminous infrared galaxies (LIRG) is difficult due to their extreme obscuration. Observations at radio and mm wavelengths can penetrate large columns of dust and gas and provide unique insights into the properties of the compact obscured nuclei of LIRGs.}
   {The aim of this study is to constrain the dynamics, structure and feeding of the compact nucleus of NGC~4418, and to reveal the nature of the main hidden power source: starburst or AGN.}
   {We obtained high spatial resolution observations of NGC~4418 at 1.4 and 5~GHz with MERLIN, and at 230 and 270 GHz with the SMA very extended array. We use the continuum morphology and flux density to estimate the size of the emitting region, the star formation rate and the dust temperature. Emission lines are used to study kinematics through position-velocity diagrams. Molecular emission is studied with population diagrams and by fitting an LTE synthetic spectrum.}
   {We detect bright 1~mm line emission from CO, HC$_3$N, HNC and C$^{34}$S and 1.4~GHz absorption from HI. The CO 2--1 emission and HI absorption can be fit by two velocity components at 2090 and 2180 km\,s$^{-1}$. We detect vibrationally excited HC$_3$N and HNC, with $T_{\rm vib}\sim$300 K. Molecular excitation is consistent with a layered temperature structure, with three main components at 80, 160 and 300 K. For the hot component we estimate a source size of less than 5 pc. The nuclear molecular gas surface density of 10$^4$ M$_\odot$ pc$^{-2}$ is extremely high and similar to that found in the ultra-luminous infrared galaxy (ULIRG) Arp220.}
   {Our observations confirm the presence of a molecular and atomic in-flow, previously suggested by Herschel observations, which is feeding the activity in the center of NGC~4418. Molecular excitation confirms the presence of a very compact, hot dusty core. If a starburst is responsible for the observed IR flux, this has to be at least as extreme as the one in the ULIRG Arp~220, with an age of 3-10~Myr and a star formation rate $>$10~M$_\odot$~yr$^{-1}$. If an AGN is present, it must be extremely Compton-thick.
}


   \maketitle
%
%________________________________________________________________

\section{Introduction}

Luminous infrared galaxies (LIRG) radiate most of their energy as thermal dust emission in the infrared (IR) and constitute the dominant population among the most luminous extragalactic objects \citep[e.g., ][]{sanders_96}. Recent observations \citep{spoon07, aalto07, costagliola11} suggest that they may play a crucial role in galaxy evolution, representing the early obscured stages of starburst galaxies and active galactic nuclei (AGN). 

The central regions of LIRGs are deeply enshrouded in large columns of gas and dust, which preclude direct investigation in the optical and IR. Observations at radio and mm/sub-mm wavelengths can penetrate these large amounts of obscuring material and provide unique insight in the properties of LIRGs. The study of molecular line emission has proven to be a key player in this field, by providing valuable information on the physical conditions of the gas, its chemistry and dynamics. In a 3~mm line survey with the IRAM EMIR broadband receiver \citep{costagliola11} we found a small sample of HC$_3$N--luminous and HCO$^+$--faint galaxies. This objects have compact ($<$100~pc) obscured ($A_\mathrm{V}>$100 Mag) nuclei, characterized by deep silicate absorption in the mid-IR and rich molecular spectra, which cannot be explained by standard chemical models \citep[e.g., ][]{spaans_05, baan08, costagliola11}. 

The peculiar spectral properties of compact obscure nuclei (hereafter CON) resemble to some degree those of Galactic hot cores, and may be tracing the early, dust-enshrouded stages of nuclear activity. Very long baseline interferometry (VLBI) observations at radio wavelengths have revealed that the IR emission from LIRGs can be powered by both starburst and AGN activity \citep[e.g., ][]{miguel09, batejat11}. LIRGs can have star formation rates (SFR) 10 to 100 times greater than what is observed in the Milky Way, and very high gas surface densities, which resemble the properties of star-forming sites in high redshift submillimeter galaxies (SMGs) \citep[e.g., ][]{swinbank11}. Whether the high SFR is due to the extreme gas surface densities \citep[e.g., ][]{kennicutt98}, or to an increased star formation efficiency in high-pressure environments \citep[e.g., ][]{elmegreen97}, is still a matter of debate. Compact obscured nuclei are ideal laboratories to study chemistry, molecular excitation and star formation in extreme interstellar medium (ISM) conditions.

\subsection{NGC~4418: The prototypical CON}

The luminous infrared galaxy (LIRG) NGC~4418 ($L_{\rm IR}=10^{11}$~L$_\odot$) has the optical morphology of an early-type spiral and was first detected as a bright source by the IRAS satellite. Lying at $D$=34~Mpc, the galaxy is part of an interacting pair, with the companion being an irregular blue galaxy roughly two arc-minutes (16~kpc) to the south-east. 

The optical spectrum of NGC~4418 has been described by \cite{roche86} as {\it unremarkable}, with only faint emission from S[II] and H$_{\rm\alpha}$ on a strong continuum, and does not betray the presence of the bright IR source. This is explained by mid-IR observations \citep{spoon_01, roche86}, which show a deep silicate absorption at 10~${\rm\mu}$m, one of the deepest ever observed in an external galaxy, corresponding to roughly 100 magnitudes of visual extinction. Whatever is powering the large IR flux of NGC~4418, it must be hidden behind a thick layer of dust, which makes it undetectable at optical wavelengths.

The large IR luminosity requires a compact starburst or an AGN to heat up a large column of dust. However, because of the high obscuration, the direct investigation of the nuclear region is a challenging task, and the nature of the energy source still unclear.  Observations with the {\it Chandra} X-ray satellite by \cite{maiolino03} show a flat hard X-ray spectrum, which would be an indication of a Compton-thick AGN, but the photon statistics is too limited to be conclusive. The absence of a clear X-ray signature may imply either that the galaxy is powered by star formation alone, or that the obscuring column is so high that most of the X-ray cannot escape its nuclear region.

Additional evidence of nuclear activity in NGC~4418 comes from near and mid-IR observations.  High-resolution {\it HST }near-infrared, and Keck mid-infrared images by \cite{evans03} show that the galaxy has a 100-200 pc optically thick core,  with an high IR surface brightness. The observed spectral energy distribution implies a dust temperature of 85 K, which, if compared to the total IR flux, implies the presence of an optically thick source of no more than 70~pc across. The existence of a  compact source is confirmed also by observations of vibrationally excited HC$_3$N \citep{costagliola2010} and HCN \citep{sakamoto10}, which reveal dust temperatures of the order of 200-500~K, and a source size of less than 10 pc. 

In a recent study \citet{sakamoto2013} have used SMA observations at 860 and 450 $\mu$m to directly probe the nucleus of NGC~4418 at sub-arcsecond resolution. This study confirms the existence of a $\sim$20~pc (0.$''$1) hot dusty core, with a bolometric luminosity of about 10$^{11}$ L$_\odot$, which accounts for most of the galaxy luminosity. The high luminosity-to-mass ratio ($L/M\simeq$500 L$_\odot$ M$_\odot^{-1}$) and luminosity surface density (10$^{8.5\pm0.5}$ L$_\odot$ pc$^{-2}$) are consistent with a Compton-thick AGN to be the main luminosity source. Alternatively, an extreme (SFR$\simeq$100 M$_\odot$ yr$^{-1}$), young ($\leq$5 Myr), compact starburst could also reproduce the properties of the inner core. By comparing SDSS optical images in the {\it i$^\prime$, z$^\prime$, g$^\prime$, r$^\prime$} filters, a reddening feature was found perpendicular to the major axis of the galaxy, which is interpreted as an outflow cone emanating from the nucleus.

NGC~4418 was first shown to have a rich molecular chemistry by \citet{nascent} and the large abundance of HC$_3$N \citep[$>$10$^{-8}$,][]{costagliola2010} is not obviously consistent with an X-ray dominated chemistry expected in an AGN scenario \citep{meijerink07}. Together with a low HCO$^+$/HCN $J$=1--0 line ratio, bright HC$_3$N is instead reminiscent of line emission towards Galactic hot-cores, i.e. regions of dense, warm, shielded gas around young stars. This has led some authors to propose that NGC~4418 may be a very young starburst, where the star forming regions are still embedded in large amounts of dust \citep{nascent, costagliola11}. This scenario of a {\it nascent} starburst would be consistent with the galaxy being radio-deficient \citep{roussel03}. However, this picture is complicated by the possibility of a buried AGN, the required extreme properties of a buried starburst, and by recent chemical model developments suggesting that substantial HC$_3$N abundances may occur near AGNs \citep{harada10}.

Recent Herschel PACS observations \citep{galfonso2012} reveal the presence of a compact (5~pc), hot (350~K) core, and red-shifted OH absorption, which is interpreted as the signature of a molecular in-flow. These observations confirm previous estimates of the properties of the compact core \citep[e.g., ][]{costagliola2010}, and suggest that a gas in-flow may be feeding the central engine. However, both single-dish and satellite observations lack the spatial resolution needed to directly probe the inner 100~pc and are thus highly model-dependent. The high angular resolution observations obtained with the MERLIN VLBI network presented here, and the Submillimeter Array (SMA) observations {\it presented here and in the companion paper \citep{sakamoto2013}}, allow us to investigate the properties of the nuclear core with unprecedented accuracy.
\begin{table}
%\begin{center}
\caption{\label{tab:props} General properties of NGC~4418 assumed in the text.}
% NGC4418 generic properties 

\renewcommand{\arraystretch}{1.2}
\setlength{\tabcolsep}{2pt}
\begin{tabular}{lcr} 
\multicolumn{3}{c}{\bf Properties of NGC~4418} \\
\hline 
\hline 
 &  &\\ 
Phase Center (RA, Dec, J2000)$^1$ & &  12:26:54.611, $-$00:52:39.42\\
Distance$^2$ & & 34$\pm$2~Mpc\\
Scale at Hubble flow distance$^2$ & & 165 pc/arcsec \\
Systemic Velocity (LSR)$^3$ & & 2090 km s$^{-1}$\\
IR Luminosity (40--120~$\mu$m)$^4$ & & 10$^{10.63}$~L$_\odot$ \\
IR Luminosity (8--1000~$\mu$m)$^5$ & & 10$^{11.1}$~L$_\odot$ \\
Classification$^2$ & &  LIRG, Seyfert 2\\
&  &\\
\hline 
\hline
\end{tabular}\\
{\tiny
$^1$ Common phase center of all observations \\ (position of the CO 2--1 intensity peak) \\
$^2$ From the NASA/IPAC Extragalactic Database (NED):\\ \url{http://ned.ipac.caltech.edu/} \\
$^3$ From HNC kinematics in \citet{sakamoto2013} \\
$^4$ From \citet{baan06}\\
$^5$ From IRAS flux in \citet{sanders03}\\
}

%\end{center}
\end{table}
\section{Observations}

\subsection{MERLIN}

NGC~4481 was observed with MERLIN\footnote{MERLIN is a national facility operated by The University of Manchester on behalf of the Science and Technology Facilities Council (STFC)}, including the 76-m Lovell Telescope, on 30th/31st May 2005. The total duration of the observations was 12.4~hrs, including 7.8~hrs on-source observation of NGC~4418. Throughout the observing run, observations of NGC~4418 were regularly interspersed with scans on the nearby phase reference calibration source 1216-010. A single scan of 3C~286 was used as the primary flux density calibrator and the bright point source calibrator 0552+398 was used for secondary flux density calibration and to derive bandpass solutions. Both hands of circular polarization were recorded over a total bandwidth of 4~MHz, centered on the red-shifted frequency of HI (1410.165~MHz). The total bandwidth of the observations was correlated into 128 channels with width of 31.25~kHz equating to a velocity resolution of 6.5~km~s$^{-1}$. Initial editing and calibration was performed at Jodrell Bank Observatory using the local MERLIN {\sc dprogs} software before the data were exported to {\sc fits}. Further calibration, including flux density, bandpass and phase reference calibration were performed using the MERLIN pipeline, following standard procedures. 

In addition to these MERLIN 1.4~GHz spectral line data, 5~GHz continuum data were extracted from the MERLIN archive. These data were observed on 3rd June 2001 as part of a combined MERLIN and EVN experiment. In total the observation lasted 7.8~hrs, with an on-source integration of 2.8~hrs on NGC~4418. Regular observations of the two phase reference calibrators J1232-0224 and J1229+0203 were made throughout the run and a single 20~min scan of 3C\,286 was used for flux density calibration. Observations of J0555+3948 and OQ208 were used to derive secondary flux calibration and band-pass. Six MERLIN telescopes (not including Lovell Telescope) were used, with the data correlated into 16 1~MHz-wide channels in all four Stokes.  These continuum data were initially edited and calibrated using local MERLIN software, before being exported to {\sc fits} for further analysis.

The calibrated data were imaged, cleaned and analyzed with the {\sc gildas/mapping}\footnote{http://www.iram.fr/IRAMFR/GILDAS} data reduction package. To easily compare the different observations, a common phase center was set at  $\alpha_\mathrm{J2000}$=12:26:54.611 and $\delta_\mathrm{J2000}$=$-$00:52:39.42, corresponding to the position of the peak of the CO 2--1 integrated intensity (see Section \ref{sec:co} and Table \ref{tab:props}). The MERLIN observations were deconvolved using a Clark clean algorithm with uniform weighting, resulting in a beam size of 0.$\!''$35$\times$0.$\!''$16 and position angle (PA) of 23$^\circ$ at 1.4~GHz, and a beam size of 0.$\!''$13$\times$0.$\!''$04 with PA 19$^\circ$ at 5~GHz. The deconvolved spectral data cube at 1.4~GHz was smoothed to a velocity resolution of 20 km\,s$^{-1}$. The spatial resolution of our observations is roughly 40 and 15 pc, for observations at 1.4~GHz and 5~GHz, respectively.

\subsection{SMA}

Observations were carried out on February 24th and 27th 2010 with the SMA \footnote{The Submillimeter Array (SMA) is a joint project between the Smithsonian Astrophysical Observatory and the Academia Sinica Institute of Astronomy and Astrophysics and is funded by the Smithsonian Institution and the Academia Sinica.} on Mauna Kea, Hawaii. All eight array antennas were available and placed in the very extended configuration, with baselines ranging 120-500 m. Observations consisted on two tracks where the SIS receivers were tuned at the rest frequencies of 230.537~GHz, and 271.981~GHz in the lower side band.

The phase reference center of the observations was $\alpha_\mathrm{J2000}$=12:26:54.60 and $\delta_\mathrm{J2000}$=$-$00:52:39.0. Gain calibration was checked every 15 min by observing the nearby ($\sim3^\circ$) quasar 3C\,273. Bandpass calibration was derived from the spectrum of J0854+201, while absolute flux calibration was done on Mars, Titan, and Ceres. Average zenith opacities at 225~GHz during the observations were of 0.07 and 0.09, corresponding to a precipitable water vapor (PWV) of $\sim 1$ and 2 mm, respectively. Data calibration and reduction was performed using the MIR-IDL package.

The calibrated data were deconvolved using a Clark CLEAN algorithm with uniform weighting, resulting in a beam size of 0.$\!''$49$\times$0.$\!''$38 and 0.$\!''$41$\times$0.$\!''$33 at 230 and 270 GHz respectively, and a position angle of 60$^\circ$. The resulting deconvolved spectral data cubes were smoothed to a common velocity resolution of 20 km\,s$^{-1}$.  The spatial resolution of our observations at 1~mm is roughly 50 pc.

 \begin{figure*}
   \centering
   \includegraphics[width=.9\textwidth,keepaspectratio]{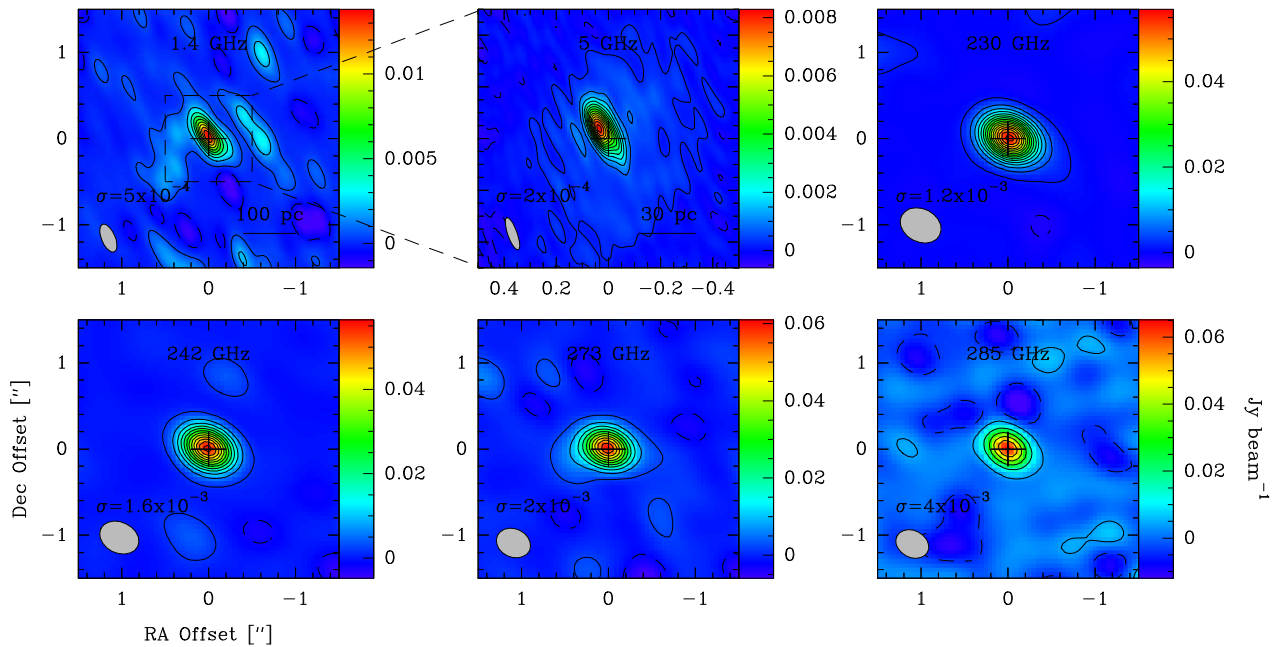}
             \caption{Maps of continuum emission for MERLIN and SMA observations in NGC~4418. Contours are drawn every 3~$\sigma$, starting at 1-$\sigma$. Negative contours are drawn as dashed lines. Position offsets are relative to the phase center reported in Table \ref{tab:props}.}
         \label{fig:cont}
   \end{figure*}

\begin{table}
\begin{center}
\caption{\label{tab:cont}Properties of continuum emission in NGC~4418. For the flux density we report the peak value in mJy/beam, and the total integrated flux density in mJy (rms values are shown in parentheses). The circle-averaged source sizes were computed via visibility-fits in the $u-v$ (see Fig. \ref{fig:sizes}), for unresolved emission we report upper limits.}
% Results Continuum 
% Freq Flux(beam) Flux(tot) beamsize sourcesize  
\renewcommand{\arraystretch}{1.2}
\setlength{\tabcolsep}{2pt}
\begin{tabular}{ccccc} 
\hline 
\hline 
& & & &\\ 
Freq.  & Flux dens. (peak, int.) & Beam & Source Size & K/Jy \\
GHz  & mJy beam$^{-1}$, mJy  & $maj\times min,$ $PA$ & $circ.$ $aver.$ &\\
\hline 
& & & &\\ 
\multicolumn{5}{l}{MERLIN:} \\
%& & & &\\ 
1.4 & 14 (0.3), 38 (2) & 0.$\!''$35$\times$0.$\!''$16, 23$^\circ$ & 0.$\!''$5, 80 pc &  1.1$\times$10$^7$\\ 
5 & 8 (0.2), 34 (2) & 0.$\!''$13$\times$0.$\!''$04, 19$^\circ$ & 0.$\!''$15, 25 pc & 1.0$\times$10$^7$\\
& & & &\\ 
\hline 
& & & &\\ 
\multicolumn{5}{l}{SMA:} \\
%& & & &\\ 
230 & 57 (1),  59 (1) & 0.$\!''$48$\times$0.$\!''$38, 62$^\circ$ & $<$0.$\!''$4, 65 pc & 121 \\
242 & 56 (1),  57 (1) & 0.$\!''$48$\times$0.$\!''$38, 62$^\circ$ & $<$0.$\!''$4, 65 pc & 123\\
273 & 61 (2),  61 (1) & 0.$\!''$41$\times$0.$\!''$33, 64$^\circ$ & $<$0.$\!''$3, 50 pc & 122\\
285 & 65 (4),  65 (1) & 0.$\!''$41$\times$0.$\!''$33, 64$^\circ$ & $<$0.$\!''$3, 50 pc & 122\\
 & & & & \\ 
\hline
\hline
\end{tabular}

\end{center}
\end{table}

 \begin{figure*}[t]
   \begin{centering}
   \includegraphics[width=.9\textwidth,keepaspectratio]{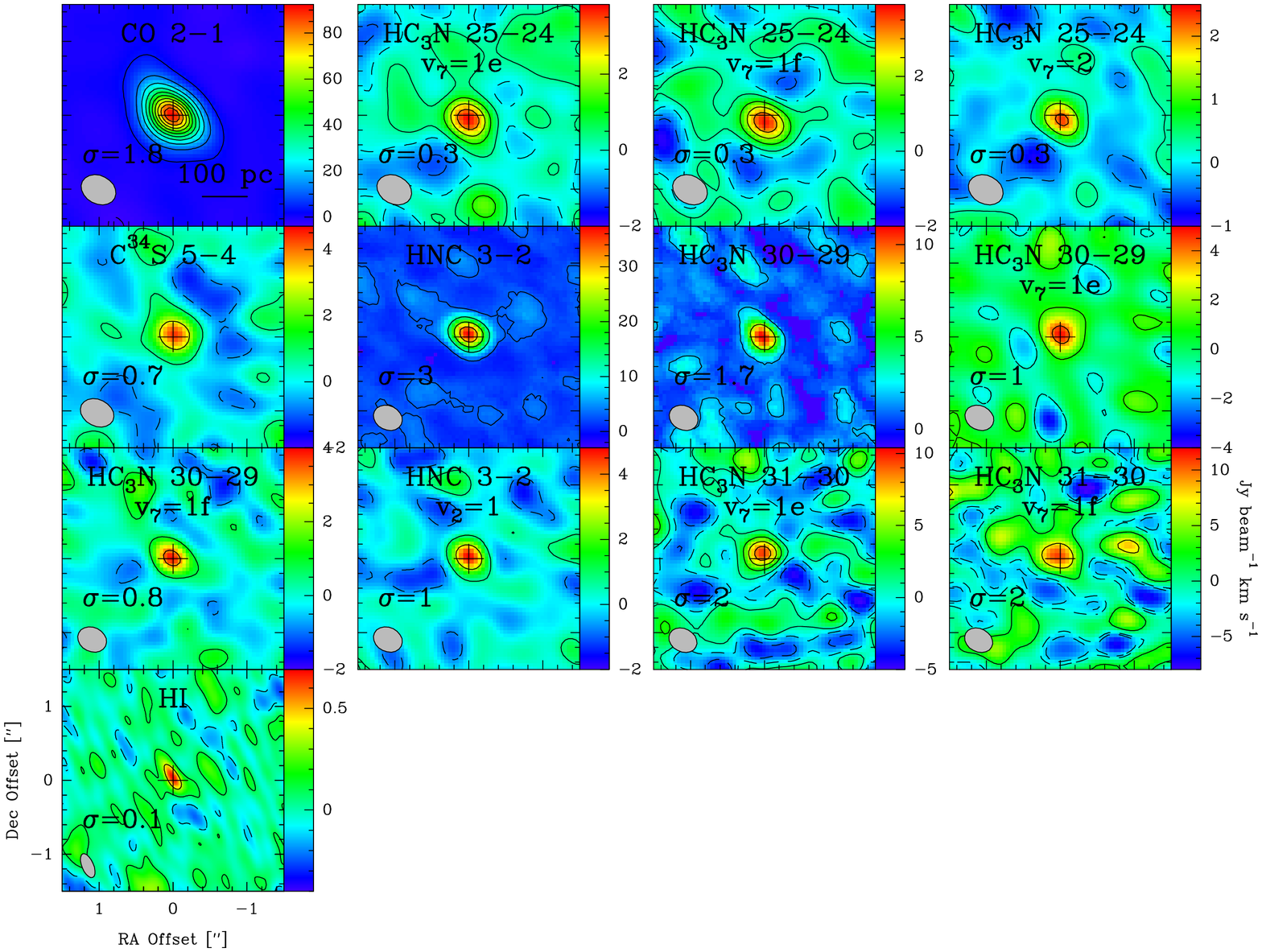}  
    \caption{Continuum-subtracted integrated intensity maps of all lines detected in the SMA and MERLIN bands. Contours are drawn every 5~$\sigma$ for CO, and every 3~$\sigma$ for the other lines, starting at  1-$\sigma$. Negative contours are drawn as dashed lines. For convenience, the intensity scale for HI absorption has been inverted. Position offsets are relative to the phase center reported in Table \ref{tab:props}.}
         \label{fig:mom0}
\end{centering}
   \end{figure*}

\section{Results}

\subsection{The 1.4~GHz and 5~GHz continuum}
\label{sec:radcont}

The maps of continuum emission are shown in Fig. \ref{fig:cont}, while the integrated intensity maps of the detected lines can be found in Fig \ref{fig:mom0}. Source sizes were derived for both continuum and spectral line emission via fitting elliptical Gaussians in the $uv$-domain, with the {\sc mapping} routine {\sc uv\_fits}, and compared to circle--averaged visibility profiles (see Fig. \ref{fig:sizes}).  

The radio continuum in NGC~4418 was detected in both MERLIN bands, with integrated total flux densities of 38 and 34 mJy at 1.4 and 5~GHz respectively. Our data are in agreement with previous observations with the VLA by \cite{baan06}. These authors detect the same total flux density in a similar beam, but do not discuss the structure of the emission, assuming the source to be unresolved. In our data, from both uv-fitting and circle-averaged profiles, the emission appears to be spatially resolved,  with source sizes of 0.$\!''$5 at 1.4~GHz and 0.$\!''$15 at 5~GHz, corresponding to a linear extent of 80 and 25 pc (see Fig. \ref{fig:cont}, \ref{fig:sizes}).

At both 1.4 and 5~GHz, filaments departing from the central peak are detected at the 3-$\sigma$ level. These filaments have an orientation roughly perpendicular to the major axis of the galaxy (from 2MASS, PA$\simeq$60$^\circ$, \citet{2mass} ) and extend up to 100 and 30 pc out of the projected galactic plane at 1.4 and 5~GHz , respectively. This detection is, however, still to be considered tentative, due to the poor beam quality of the MERLIN interferometer at low declinations, and has to be confirmed.

\subsection{HI absorption}
\label{sec:h1}
The 21 cm atomic hydrogen line is clearly detected in absorption towards the nucleus of NCG~4418. The HI absorption is spatially resolved, covering the whole extent of the 1.4~GHz continuum. We estimate a lower limit for the angular diameter of the absorbing HI gas of 0.$\!''$5, corresponding to a linear size of 70 pc at the galaxy distance. The maximum of the absorption is slightly displaced from the phase center, and from the maximum of the 1.4~GHz continuum, with offsets +0.$\!''$05 and $-$0.$\!''$05 in R.A. and Dec., respectively. 

\begin{figure*}[t]
   \centering
   \includegraphics[width=0.9\textwidth,keepaspectratio]{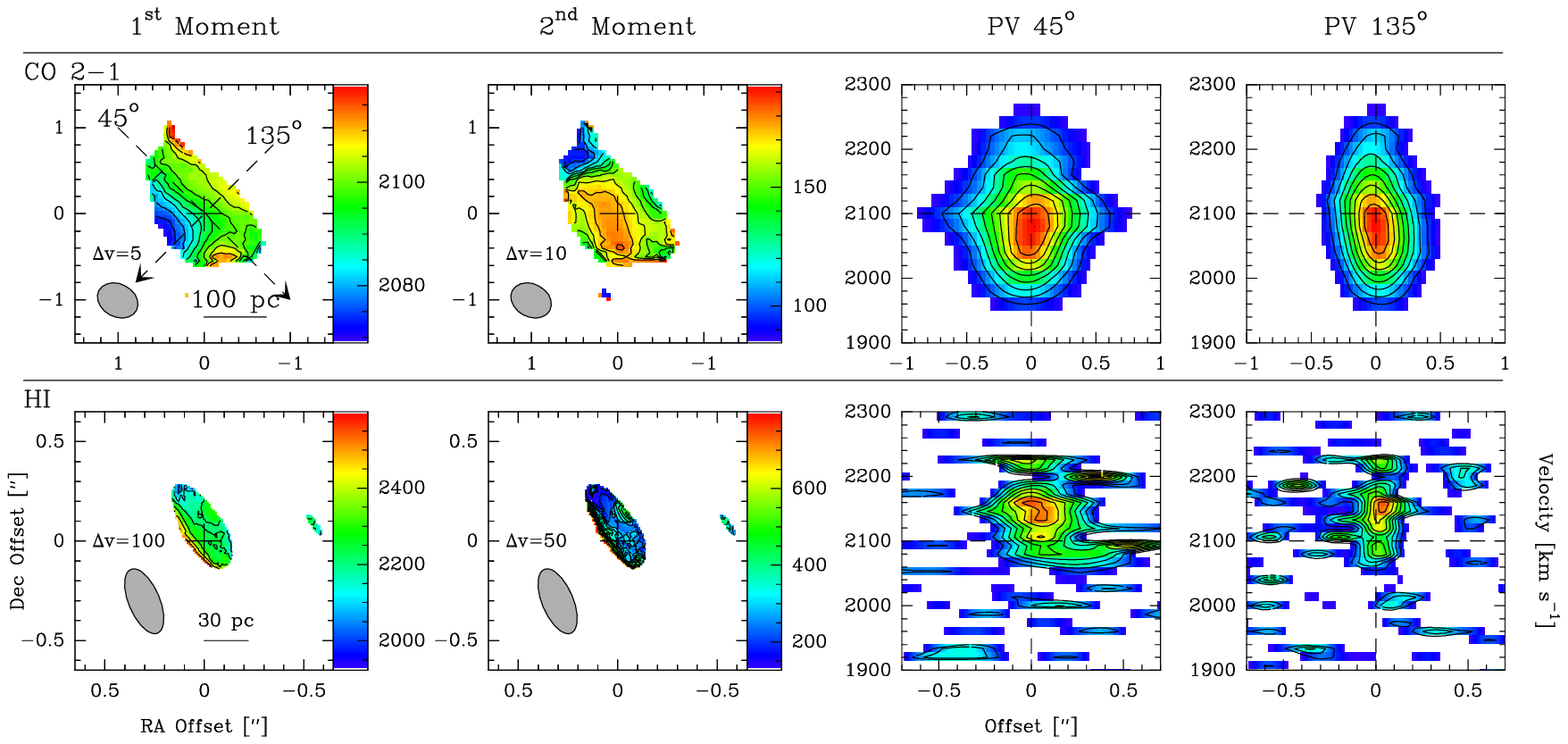}
            \caption{Kinematics of CO~2--1 emission and HI absorption. The first and second moments of the deconvolved data-cubes are shown, together with position-velocity diagrams along the major (45$^\circ$) and minor (135$^\circ$) axis of the molecular emission. The color scale starts at a signal-to-noise ratio of 5. In the first and second moment maps, the contour spacing is shown as $\Delta$v, in km\,s$^{-1}$. Notice the velocity shift in the PV diagrams of CO~2--1 and HI. For discussion, see Sections \ref{sec:h1}, and \ref{sec:co}.}
         \label{fig:dynamics}
   \end{figure*}

In Fig. \ref{fig:dynamics} we show the first and second moment maps of HI, together with position-velocity (PV) diagrams along PA=45$^\circ$ and PA=135$^\circ$. As projection angles we choose the major and minor axis of the integrated CO~2--1 emission (see Figs. \ref{fig:mom0} and \ref{fig:dynamics}). The velocity-centroid map and the PV diagrams show no clear sign of rotation, however, a velocity gradient of roughly 70 km\,s$^{-1}$ pc$^{-1}$ is marginally detected along the major axis of the galaxy. 

The continuum-subtracted HI spectrum, extracted from the central beam, is shown in Fig. \ref{fig:spec}(c). The line profile appears skewed, with a marked blue excess extending to LSR velocities lower than 2000 km\,s$^{-1}$. The asymmetric line is well fit by two Gaussian velocity components, with centers at 2150 and 2090 km\,s$^{-1}$, and FWHM line widths of 134 and 160 km\,s$^{-1}$, respectively. The results of the fit are shown in Table~\ref{tab:lines}.

\subsection{The 1 mm continuum}
\label{sec:mmcont}

The 1~mm continuum was extracted from the line-free channels in each side band of the SMA observations. Maps of the continuum emission in the four bands are shown in Fig. \ref{fig:cont}. 

A circle-averaged source size was calculated from visibility fits (see Fig. \ref{fig:sizes}), the results are reported in Table~\ref{tab:cont}. The emission is unresolved in all 1~mm bands, with upper limits for the angular size of the emission of 0.$\!''$4 for the LSB and USB of the 230~GHz tuning and 0.$\!''$3  for both sidebands of the 270~GHz tuning, corresponding to linear extents of 60 and 45~pc. Total flux densities range from 59 mJy at 230 GHz to 65 mJy at 285~GHz. Given these results, we assume a 1~mm flux density of 60 mJy confined in a region smaller than 50 pc in diameter. The Jansky-to-Kelvin conversion factor for our SMA observations is roughly 122, which gives a lower limit for the brightness temperature of the 1~mm continuum of 70 K. 
The presence of a compact 1-mm core is consistent with earlier SMA observations \citep{sakamoto2013}, where we find part of the continuum at 860 and 450 $\mu$m to be emerging from a region of less than 0$''$.1 in diameter, with brightness temperature greater than 100~K.

\subsection{Molecular Emission}
\label{sec:smamol}
\begin{table*}
\begin{center}
\caption{\label{tab:lines} Properties of spectral lines detected in NGC~4418. Results of Gaussian fitting are shown, together with circle-averaged source sizes. Line peak flux densities refer to the flux density integrated on the central clean beam. The conversion from flux density to main beam brightness $T_\mathrm{mb}$  is done by applying the K/Jy conversion factor of Table~\ref{tab:cont}. For the resolved emission, we report the deconvolved brightness temperature $T_\mathrm{b}$, or its lower limit in the case of unresolved emission. Errors on the measured flux densities and brightness temperatures are reported in parentheses. Line-center velocities V$_{\rm c}$ are LSR optical velocities. For HNC 3-2,  for which two alternative fits are possible, these are labeled with $(a)$ and $(b)$ (see Section \ref{sec:hnc}). Source sizes are full-width-half-maximum sizes derived from fitting circle-averaged visibilities with circular Gaussian profiles (see Fig. \ref{fig:sizes}). In the presence of strong line blending, the velocity and line width of the emission were fixed to the values derived for blend-free transitions. Fixed values are marked with an asterisk in the table.}
% Results Lines 
% line freq flux(tot) Tb vc dv
%\renewcommand{\arraystretch}{1.2}
%\setlength{\tabcolsep}{2pt}
%\rowcolors {4}{gray!35}{}
%\begin{tabular}{lllllllll} 
\begin{tabular}{llccccccc} 
\hline 
\hline 
& & & & & & & &\\ 
& Line & Rest Frequency & E$_\mathrm{u}$ & Peak Flux & T$_\mathrm{b}$ & v$_\mathrm{c}$ & $\Delta$v & Source Size\\
&  & [GHz] & [K] & [mJy] & [K] & [km s$^{-1}$] & [km s$^{-1}$] & $circ.$ $aver.$ \\
\hline 
& & & & & & & &\\ 
& HI & 1.420 & 0.07 & -6 $\pm$ 0.5 & - & 2153 $\pm$ 11 & 131  $\pm$ 12  & $>$0.5 $''$, 80 pc \\
&    & $''$ & $''$ & -2 $\pm$ 0.5 & - & 2094 $\pm$ 22 & 160 $\pm$ 18  & $>$0.5 $''$, 80 pc \\
& & & & & & & &\\ 
& CO 2-1 & 230.538 & 17 & 439  $\pm$ 6 & 80  $\pm$ 0.6 & 2091 $\pm$ 3 & 163 $\pm$ 6  & 0.7 $''$, 120 pc \\
&        & $''$ & $''$ & 140  $\pm$ 6 & 26  $\pm$ 0.6 & 2182 $\pm$ 13 & 165 $\pm$ 11  & 0.7 $''$, 120 pc \\
& & & & & & & &\\ 
& HC$_3$N 25-24, v$_6$=1e & 227.793 & 860 & 9  $\pm$ 6 & $>$ 2.2  $\pm$ 1.4  & 2120* & 100* & $<$0.4 $''$, 65 pc \\
& HC$_3$N 25-24, v$_6$=1f & 227.970 & 860 & 9  $\pm$ 6 & $>$  2.2  $\pm$ 1.4 & 2122 $\pm$ 9  & 93.8 $\pm$ 19  & $<$0.4 $''$, 65 pc   \\
& HC$_3$N 25-24, v$_7$=1e & 227.977 & 463 & 36  $\pm$ 6 & $>$  8.8  $\pm$ 1.4  & 2120*  & 100*  & $<$0.4 $''$, 65 pc \\
& HC$_3$N 25-24, v$_7$=1f & 228.303 & 463 & 36  $\pm$ 6 &  $>$ 8.8  $\pm$ 1.4  & 2120*  & 100*  & $<$0.4 $''$, 65 pc  \\
& HC$_3$N 25-24, v$_7$=2 & 228.858 & 787 & 12  $\pm$ 6 &  $>$ 3.0  $\pm$ 1.4  & 2120*  & 100*  & $<$0.4 $''$, 65 pc  \\
& & & & & & & &\\ 
& C$^{34}$S 5-4 & 241.016  & 28 & 43  $\pm$ 3 & 13  $\pm$ 1 & 2143 $\pm$ 4  & 77  $\pm$ 11  & 0.35 $''$, 57 pc \\
& & & & & & & &\\ 
\ldelim\{{2}{10mm}[HNC (a)] & HNC 3-2 & 271.981 & 26 & 167  $\pm$ 7 & $>$  40.6  $\pm$ 1.6 & 2090*  & 130 $\pm$ 17  & $<$0.3 $''$, 50 pc \\
\vspace{1mm}
	    &   & $''$  & $''$ & 54  $\pm$ 7 & $>$  13.2  $\pm$ 1.6 & 2180*  & 141 $\pm$ 21  & $<$0.3 $''$, 50 pc  \\
\ldelim\{{2}{10mm}[HNC (b)] & HNC 3-2 & 271.981  & 26 &   167  $\pm$ 7 & $>$ 40.6  $\pm$ 1.6 & 2100*  & 130 $\pm$ 17  & $<$0.3 $''$, 50 pc \\
& HNC, v$_2$=1e & 271.924  & 692 & 43  $\pm$ 7 &  $>$ 10.4  $\pm$ 1.6 & 2100*  & 85 $\pm$ 21  & $<$0.3 $''$, 50 pc \\
& & & & & & & &\\ 
& HC$_3$N 30-29 & 272.884 & 203 & 79  $\pm$ 7 & $>$  19.2  $\pm$ 1.6 & 2140  $\pm$ 11  & 115  $\pm$ 6  & $<$0.3 $''$, 50 pc \\
% HC$_3$N (a)& HC$_3$N 30-29 & 272.884 &  & 79 (7) & $>$10 & 2112 & 125 & $<$0.3 $''$, 40 pc \\
%\ldelim\{{2}{5mm}[HC$_3$N (b)] & HC$_3$N 30-29 & 272.884 &  & 66 (7) & $>$8 & 2068 & 50 & $<$0.3 $''$, 40 pc \\
%& 		    & 272.884  & & 82 (7) & $>$10 & 2136 & 71 & $<$0.3 $''$, 40 pc \\
& & & & & & & &\\ 
& HC$_3$N 30-29, v$_6$=1e & 273.331 & 921 & 13  $\pm$ 7 &  $>$ 3.2  $\pm$ 1.6 & 2120*  & 80*  & $<$0.3 $''$, 50 pc \\
& HC$_3$N 30-29, v$_6$=1f & 273.546 & 921 & 13  $\pm$ 7 & $>$  3.2  $\pm$ 1.6 & 2120*  & 80*  & $<$0.3 $''$, 50 pc \\
& HC$_3$N 30-29, v$_7$=1e & 273.553 & 524 & 40  $\pm$ 7 & $>$  9.8  $\pm$ 1.6 & 2120*  & 80*  & $<$0.3 $''$, 50 pc \\
& HC$_3$N 30-29, v$_7$=1f & 273.944 & 524 & 40  $\pm$ 7 & $>$  9.8  $\pm$ 1.6 & 2120*  & 78 $\pm$ 12  & $<$0.3 $''$, 50 pc \\
& & & & & & & &\\ 
& HNC 3-2, v$_2$=1f & 273.869 & 692 & 55 $\pm$ 7 &  $>$ 13.4 $\pm$ 1.6 & 2100*  & 78  $\pm$ 11  & $<$0.3 $''$, 50 pc \\
& & & & & & & &\\ 
& HC$_3$N 31-30, v$_6$=1f & 282.660 & 934 & 13  $\pm$ 10 & $>$ 3.2  $\pm$ 2.4 & 2100*  & 80*  & $<$0.3 $''$, 50 pc  \\
& HC$_3$N 31-30, v$_7$=1e & 282.668 & 538 & 39  $\pm$ 10 &  $>$ 9.4  $\pm$ 2.4 & 2100*  & 80*  & $<$0.3 $''$, 50 pc \\
& HC$_3$N 31-30, v$_7$=1f & 283.072 & 538 & 39  $\pm$ 10 & $>$  9.4  $\pm$ 2.4 & 2112 $\pm$ 11  & 82 $\pm$ 25  & $<$0.3 $''$, 50 pc \\
& & & & & & & &\\ 
\hline
\hline
\end{tabular}

\end{center}
\end{table*}

In Fig. \ref{fig:spec} we show the spectra in the SMA bands, extracted from the central beam and continuum-subtracted. The identification of molecular emission lines was done by comparing the observed spectrum with the molecular line database {\it Splatalogue}\footnote{\url{http://www.splatalogue.net/}}, which includes transitions from the CDMS \citep{cdms},  JPL  \citep{jpl} and Lovas/NIST\footnote{\url{http://www.nist.gov/pml/data/micro/index.cfm}} spectral line catalogues, and the Spectral Line Atlas of Interstellar Molecules (SLAIM)\footnote{available at \url{http://www.splatalogue.net/}}.  The small line widths found in NGC~4418 ($\Delta$v$\simeq$100 km\,s$^{-1}$) limit line blending, facilitating the identification of molecular transitions. Where line blending is present, we derive line parameters via multi-component Gaussian fitting and by comparing our observations with a synthetic LTE spectrum (see Section \ref{sec:lte} for discussion). 

We detect emission lines of several molecular species, including CO, C$^{34}$S, HNC and HC$_3$N, with typical rms of 5 mJy beam$^{-1}$ at 230 GHz, and 10 mJy beam$^{-1}$ at 270 GHz. The integrated intensity maps of the detected molecular transitions are shown in Fig. \ref{fig:mom0}, while in Table~\ref{tab:lines} we report the results from Gaussian fitting of the line profiles. 

\subsubsection{CO $J$=2-1: } Bright CO $J$=2-1 emission was detected in NGC~4418. The emission is spatially resolved (see Fig. \ref{fig:sizes}), and the integrated intensity can be fit well by an elliptical Gaussian with major axis 0.$\!''$75  and minor axis 0.$\!''$45, corresponding to 120 and 70 pc at the galaxy distance, and a position angle of 45$^\circ$. The spectral line profile is skewed toward high velocities and is fit well by two Gaussian components, with local standard at rest velocities (LSR) of 2090 and 2180 km\,s$^{-1}$ (see Section \ref{sec:co} for further discussion) and full width half maximum (FWHM) line widths of 160 km\,s$^{-1}$. We estimate a deconvolved brightness temperature of 80 and 26 K,  respectively for the low velocity and high velocity component.

The first moment map of CO~2--1 (see Fig. \ref{fig:dynamics}) shows a small but well defined velocity gradient, perpendicular to the major axis of the galaxy. This velocity shift is also revealed by the position-velocity diagram at  135$^\circ$ in Fig. \ref{fig:dynamics}, as a slight asymmetry of roughly 500 m s$^{-1}$ pc$^{-1}$ in the SE-NW direction. In the PV diagram along the major axis (45$^\circ$), we find no clear sign of rotation. The only noticeable feature is a broadening of the CO line towards the center of the Galaxy, which is also evident from the second moment map in Fig.\ref{fig:dynamics}. Further discussion on the CO~2--1 dynamics is found in Section \ref{sec:co}.

\subsubsection{HNC $J$=3-2:}\label{sec:hnc} The HNC $J$=3-2 emission line was detected in the 270 GHz SMA band. Both 2D-Gaussian fitting in the image domain, and visibility fitting in the $u-v$, show that the HNC-emitting region in marginally resolved, with an angular diameter between 0.$\!''$2 and 0.$\!''$3, corresponding to 30--50 pc at the galaxy distance  (see \ref{fig:sizes}). 

\begin{figure}
   \begin{centering}
   \includegraphics[height=.5\textwidth,keepaspectratio, angle=-90]{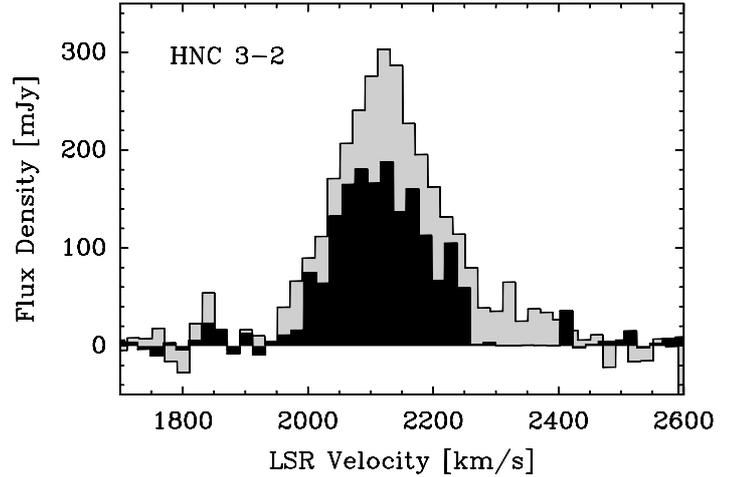}
      \caption{\label{fig:hnc} Comparison of HNC~3--2 emission observed by the IRAM 30~m telescope (grey) and with SMA (black).}
\end{centering}
   \end{figure}

The line profile appears skewed, with an evident excess at high velocities (see Figs. \ref{fig:hnc} and \ref{fig:spec}f). Two possible scenarios, described in Table~\ref{tab:lines}, fit the observed HNC emission: $a)$ two velocity components at 2090 km\,s$^{-1}$ and 2180 km\,s$^{-1}$, corresponding to the ones found for the CO~2--1 emission; $b)$ a blend with the $J$=3-2 transition of the v$_2$=1e vibrationally excited state of HNC. 

This last scenario is supported by the detection of the $J$=3-2, v$_2$=1f transition of HNC at 273.869 GHz (see Fig. \ref{fig:spec}). The line is partially blended with $J$=30-29 emission of vibrationally excited HC$_3$N, but the two lines are far enough in velocity to be clearly separated. The strength of the  v$_2$=1f line is consistent with the HNC $J$=3--2, v=0 line being blended with v$_2$=1e emission. {\it To our knowledge, this is the first extragalactic detection of mm-wave emission from vibrationally excited HNC.} In Fig. \ref{fig:spec} we only show the fit relative this second scenario, however we cannot exclude the HNC 3-2 line to receive a significant contribution from the high velocity component at 2180 km\,s$^{-1}$. Most probably, both vibrational excitation and a second velocity component concur in shaping the observed profile.

The peak flux density of the HNC $J$=3--2 transition detected by the SMA is roughly 70$\%$ of the flux density obtained by single dish observations with the IRAM~30~m telescope (Fig. \ref{fig:hnc}). Our observations were done with the extended configuration of the SMA, with a minimum projected baseline of 80 m, which converts into a maximum detectable source size of 3 $''$. The missing flux may be thus coming from an extended component ($>$1.$\!''$5) which is resolved out by our SMA observations. Such an interpretation is also suggested by the line profile, which is wider for the SMA spectrum, as we would expect if an extended, narrow component had been filtered out because of the incomplete uv-coverage.

\subsubsection{HC$_3$N:}
 \begin{figure*}[h!t]
   \begin{centering}
   \includegraphics[width=.9\textwidth,keepaspectratio]{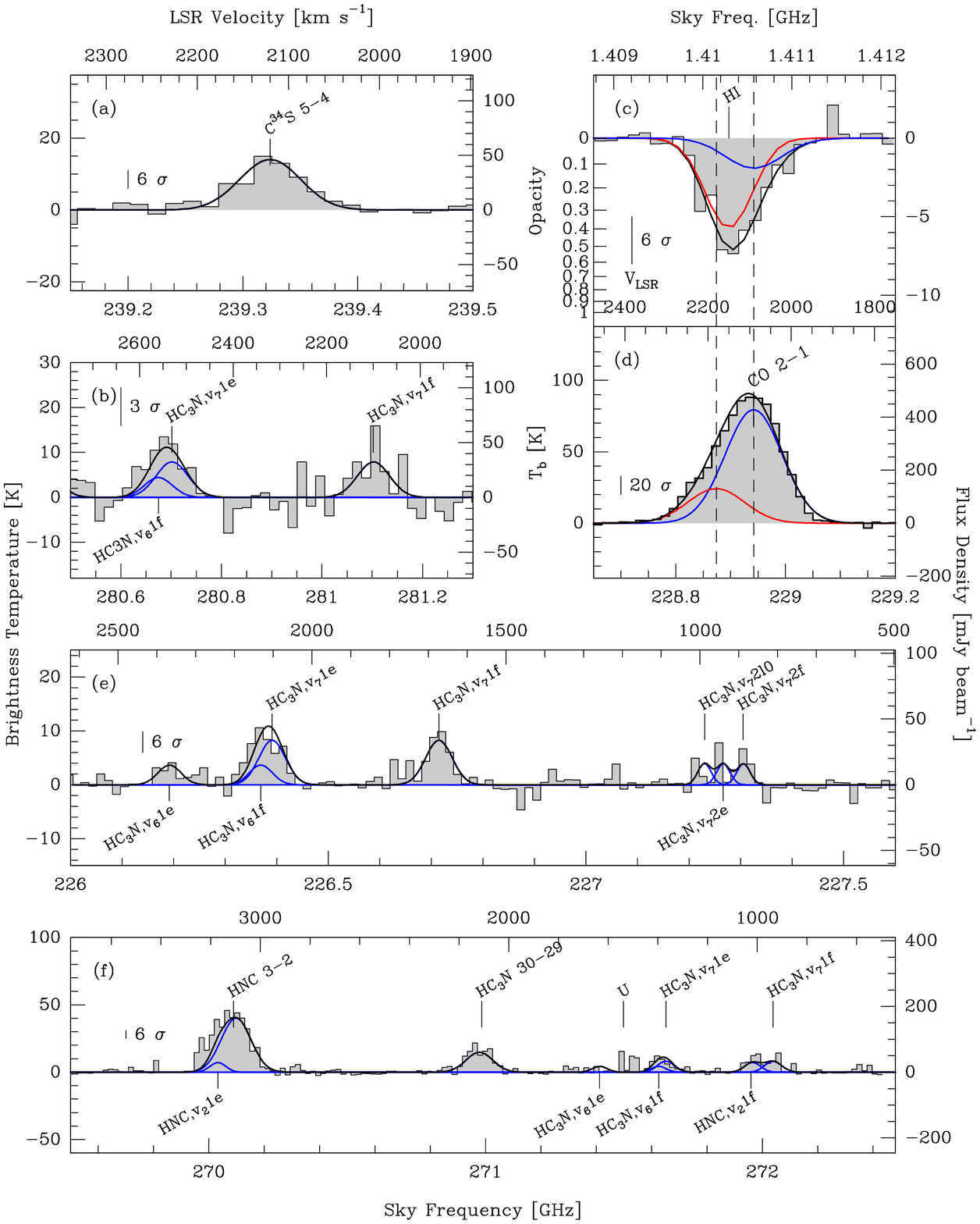}
      \caption{Spectra extracted from the central beam of MERLIN and SMA observations in NGC~4418. The data are drawn as shaded histogram, the solid black line represent the total profile of our LTE best-fit model, and LTE emission from single molecular species is drawn in blue. Panels $(c)$ and $(d)$ show a comparison of HI and CO~2--1 profiles. Dashed vertical lines mark the velocity centroid of the two components of CO emission, SC at 2090 and RC at 2180 km\,s$^{-1}$ LSR (see Section \ref{sec:co}). The red-shifted Gaussian components (RC), associated with a possible in-flow, are drawn in red in both panels. The conversion between flux density and brightness temperature is discussed in Section \ref{sec:lte}. The results of the Gaussian fit for each line are shown in Table~\ref{tab:lines}.}
         \label{fig:spec}
\end{centering}
   \end{figure*}

The HC$_3$N molecule is detected in both in the vibrational ground state and in the v$_7$=1 and v$_6$=1 excited levels (see Fig. \ref{fig:spec}). The emission is unresolved at all observed wavelengths (see Fig. \ref{fig:sizes}), which gives an upper limit for the diameter of the HC$_3$N-emitting region of about 50 pc (0.$\!''$3).  

The brightest HC$_3$N emission ($\sim$80 mJy, $T_{\rm b}\simeq$ 20 K) comes from the $J$=30--29 transition of the ground vibrational state (v=0) at 272.884 GHz. A Gaussian fit of the line profile reveals an emission velocity of 2110 km\,s$^{-1}$ and a FWHM dispersion of 125 km  s$^{-1}$ (see Table~\ref{tab:lines}).

The v$_7$=1 vibrationally excited state of HC$_3$N is detected in the $J$=25--24, $J$=30--29 and $J$=31--30 rotational transitions, with a peak flux density which is roughly half the $J$=30--29, v=0 flux density in all bands. The  v$_7$=1 and v$_6$=1 states are doublets, with single states labeled $e$ and $f$, depending on the parity of the wave function \citep{yamada86}. The v$_7$=1$e$ and  v$_6$=1$f$ lines are only $\sim$10 MHz apart and are often blended together. For this reason, the line properties of v$_7$=1 and v$_6$=1 emission were derived by Gaussian fitting of the blend-free lines, i.e. the v$_7$=1$f$ and v$_6$=1$e$ transitions. 

Gaussian fitting shows that v$_7$=1 and v$_6$=1 transitions have comparable velocity centroids (v$_c\simeq$2100 km\,s$^{-1}$) and line dispersions ($\Delta$v$\simeq$100 km\,s$^{-1}$), with the peak intensity of the v$_6$=1 emission being roughly half the v$_7$=1.

A tentative detection, 3-$\sigma$ in integrated flux density, of the $J$=25--24, v$_7$=2 triplet of HC$_3$N at 228 GHz is also reported in Table~\ref{tab:lines} and Fig. \ref{fig:spec}e. For a discussion of the vibrational excitation of HC$_3$N in NGC~4418 see Section \ref{sec:vibpump}. 

\subsubsection{C$^{34}$S:}

The J=5-4 rotational transition of C$^{34}$S at 241 GHz is clearly detected in our SMA spectrum (see Fig. \ref{fig:spec}). The emission is marginally resolved, as results from 2D Gaussian fitting and $u-v$ visibility fits (see Fig. \ref{fig:sizes}), with a source size of 0.$\!''$35, corresponding to $\sim$57~pc in the source plane. A Gaussian fit of the line profile reveals an emission velocity of 2120 km\,s$^{-1}$ and a line width of 77 km\,s$^{-1}$ (see Table~\ref{tab:lines}).  
%{\bf more bla bla here}

\section{Discussion}

\subsection{Gas dynamics and kinematics}

\subsubsection{CO~2--1 Modeling}
\label{sec:co}

 \begin{figure*}[t]
   \centering
   \includegraphics[width=0.3\textwidth,keepaspectratio]{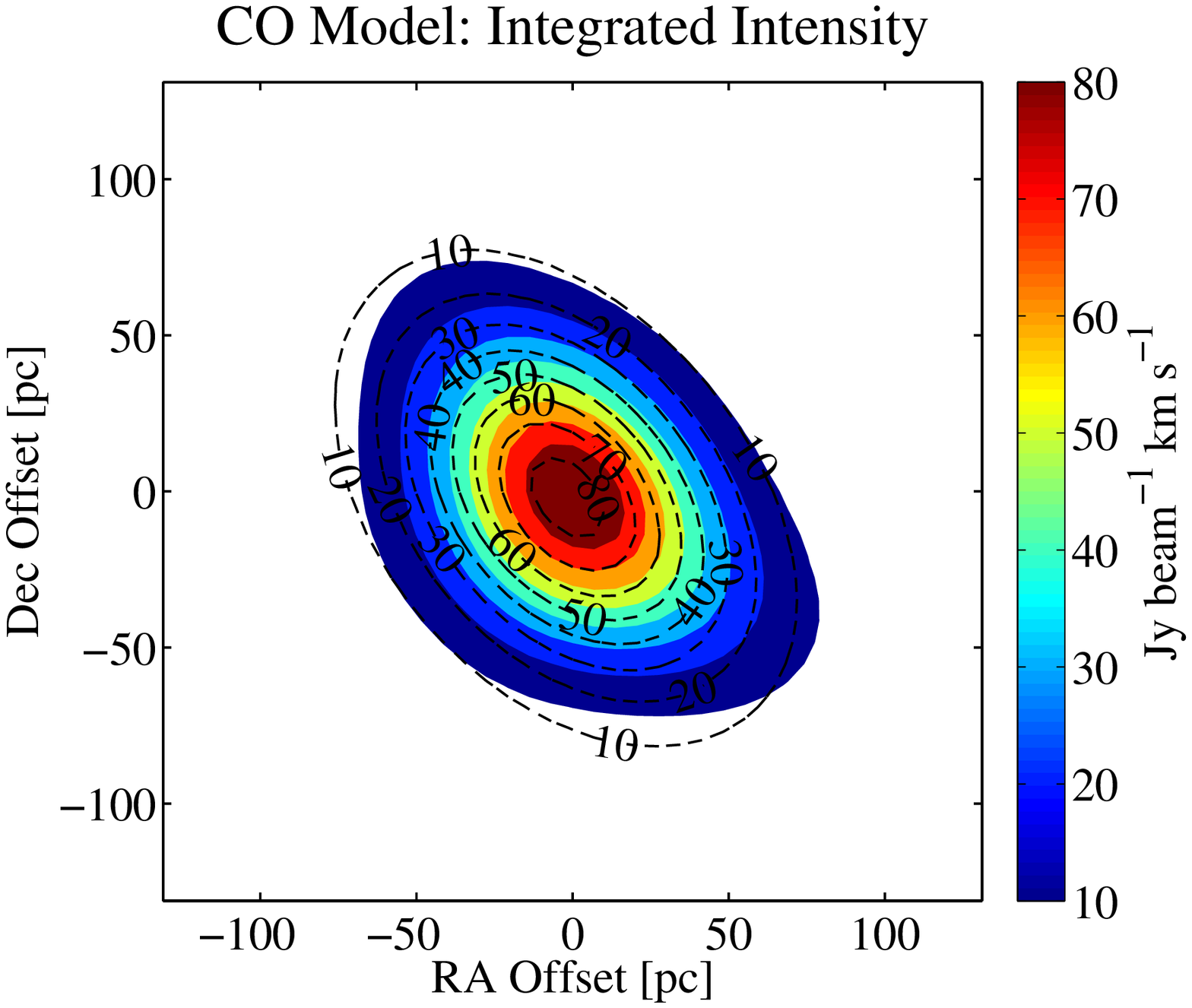}\hspace{.5cm}
   \includegraphics[width=0.31\textwidth,keepaspectratio]{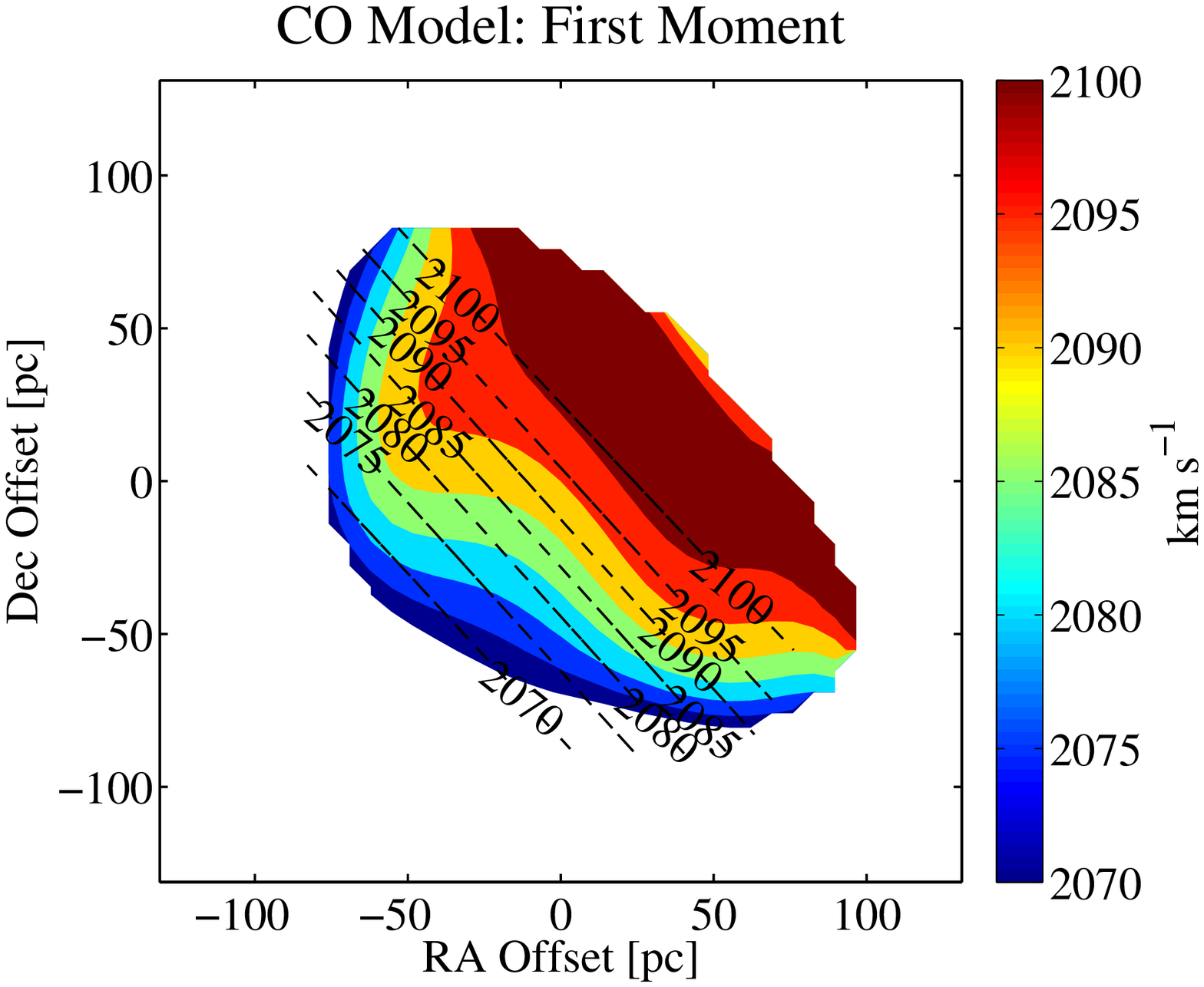}\hspace{.5cm}
   \includegraphics[width=0.3\textwidth,keepaspectratio]{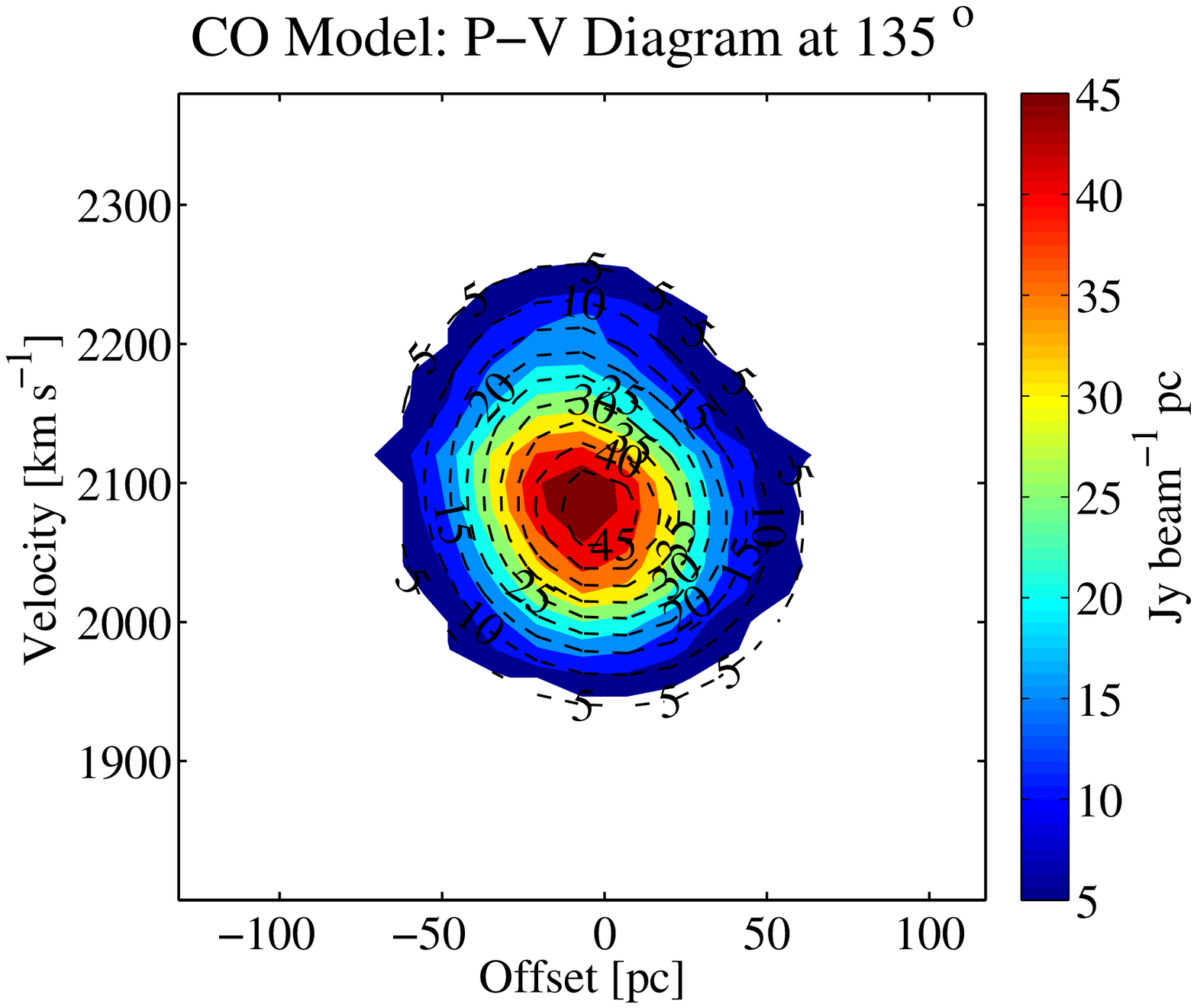}
    \caption{Model of the CO~2--1 emission. The model is overlaid as dashed contours over the color map of the observed data. The parameters of the model are reported in Table \ref{tab:model}, for discussion see Section \ref{sec:co}.}
   \label{fig:comodel}
 \end{figure*}

\begin{table}
\caption{\label{tab:model} Parameters for the two-components model for CO~2--1 emission. The model is described in Section \ref{sec:co}, while the mass estimates are discussed in Section \ref{sec:in-flow}.}
% Model Parameters
\renewcommand{\arraystretch}{1.2}
%\setlength{\tabcolsep}{2pt}
%\begin{center}
%{\bf Dynamical Model}\\
%\end{center}
\begin{tabular}{lcc}
\hline 
\hline 
& Component 1 (SC) & Component 2 (RC)\\
\hline 
& & \\
(Ra., Dec.) $^1$ & 0.03 $''$, -0.03 $''$  & -0.05 $''$, 0.05 $''$\\
(maj, min, PA) $^2$ & 0.75 $''$, 0.49 $''$, 45$^\circ$ &  0.75 $''$, 0.4 $''$, 45$^\circ$ \\
Peak $^3$ & 440 mJy &  140 mJy\\
(V$\rm_c$, $\Delta V$) $^4$ & 2090, 160 km s$^{-1}$  &  2180, 160 km s$^{-1}$\\
%(dV/dx, PA) $^5$ & -20 km s$^{-1}$ asec$^{-1}$, 135$^\circ$ & 50 km s$^{-1}$ asec$^{-1}$, 135$^\circ$\\
M(H$_2$)$^5$ & 0.6-2.5$\times$10$^8$ M$_\odot$ & 1-4.2$\times$10$^7$ M$_\odot$ \\
M(H)$^6$ & 4.9$\pm$2.9$\times$10$^5$ M$_\odot$ & 1.2$\pm$0.2$\times$10$^6$ M$_\odot$ \\
M(H$_2$)+M(H)$^7$ & 0.6-2.5$\times$10$^8$ M$_\odot$ & 1.1-4.3$\times$10$^7$ M$_\odot$ \\
 & &  \\ 
\hline
\hline
\end{tabular}
{\tiny

$^1$ Offsets of Gaussian component from phase center\\
$^2$ Major axis, minor axis and position angle of 2D Gaussian\\
$^3$ Peak intensity of CO line\\
$^4$ Velocity and FWHM of CO line\\
%$^5$ Velocity gradient and its position angle\\
%$^6$ Molecular mass calculated from CO intensity\\ 
%$^7$ Atomic mass calculated from HI opacity\\ 
%$^8$ Molecular mass ratio\\ 
$^5$ Molecular mass calculated from CO intensity (See Section \ref{sec:mass})\\ 
$^6$ Atomic mass calculated from HI opacity (See Section \ref{sec:mass})\\ 
$^7$ Total hydrogen mass\\ 
The CO 2-1 model and the derived gas masses are discussed in Section \ref{sec:co} and  Section \ref{sec:mass}, respectively.
}

\end{table}

By comparing the PV diagrams of HI and CO, it is clear that the two lines have different velocities, with the peak of HI absorption being red-shifted by roughly 60 km\,s$^{-1}$ from the peak of CO emission. This shift is confirmed by comparing the spectra extracted from the central clean beam for the two transitions, which are shown in Fig. \ref{fig:spec}(c,d). The CO profile is asymmetric, and is best fit by two Gaussian components: a 440 mJy component at 2090 km\,s$^{-1}$ and a red-shifted 140~mJy component at 2180 km\,s$^{-1}$, both with FWHM line widths of 160~km\,s$^{-1}$ (see Table~\ref{tab:lines} for details). HI absorption and CO emission show specular line profiles: both show a systemic component (SC) at 2090~km\,s$^{-1}$ and a red-shifted component (RC), but while the SC is brightest of the two for CO, the HI absorption is dominated by the RC. 

Notice that, in absence of a clear detection of rotation in the CO data cube, the systemic velocity of the galaxy is poorly constrained and the choice of the 2090 km\,s$^{-1}$ component as {\it systemic} was made only because of its brightness in the CO spectrum. However, the velocity of the SC component is consistent with the peak velocities of single dish spectra \citep[e.g.,][]{nascent, costagliola2010} and with observations of HCN 4--3, reported in \citet{sakamoto2013}, which show a velocity gradient along the major axis of the galaxy. If HCN is tracing the rotation of the inner disk, these observations set the systemic velocity of NGC~4418 at 2088 km  s$^{-1}$, which coincides with our SC component.

In order to study the kinematics traced by the CO~2--1 emission, we built a simple model, composed of two elliptical 3D Gaussian components (with coordinates:  R.A., Dec., V$_{\rm LSR}$), with the aim of reproducing the velocity gradient observed in the first moment map and PV diagram, and the integrated intensity profile.  Each component is characterized by its position on the sky (center of the ellipse), its dimensions (major axis, minor axis, position angle), the peak intensity, and by its kinematics (LSR velocity and line width). The size and position of the two components were found by fitting the channel-averaged visibilities with the {\sc uv\_fit} routine of the {\sc mapping} software. A least-square minimization in the image domain was then used to derive LSR velocities and line widths. We assume the two components to be far enough in velocity to be considered independent, as far as radiative transfer is concerned. We therefore do not include radiative transfer in our model. This is further discussed in Section \ref{sec:lte}.

The parameters that fit the data-cube are reported in Table~\ref{tab:model}. The model is consistent with two main velocity components, a systemic component at 2090 km\,s$^{-1}$ (SC) and red-shifted emission at 2180 km\,s$^{-1}$ (RC). The two components have similar sizes and orientation, but appear to be slightly displaced, with an offset of about 0.$\!''$1 in the SE-NW direction. This offset between the two components is responsible for the observed velocity gradient at PA 135$^\circ$. This simple model can reproduce 90$\%$ of the observed CO emission, a comparison with the data is shown in Fig. \ref{fig:comodel}. 

\subsubsection{An atomic and molecular in-flow}
\label{sec:in-flow}
In Sections \ref{sec:h1} and \ref{sec:co} we show that HI absorption and CO~2--1 emission can be fit by two velocity components at 2090 (SC) and 2150-2180 (RC) km\,s$^{-1}$. If we assume the SC component to be the systemic velocity of the galaxy (see discussion in Section \ref{sec:co}), the RC component observed in HI absorption is tracing red-shifted foreground gas moving towards the center of the galaxy, and it may therefore interpreted as an in-flow signature. If this scenario, the SC component in the CO 2--1 spectrum could be tracing a molecular component of the same in-flowing gas. This interpretation assumes the RC components of HI and CO 2--1 to be roughly co-spatial, which is a likely assumption given the similar velocities in the two tracers. However, the sensitivity of our HI data is not good enough to reliably constrain the morphology of the red-shifted absorption, so we cannot rule out other possible interpretations.

A spherical molecular in-flow in the core o NGC~4418 was first suggested by \citet{galfonso2012} to explain the red-shifted OH absorption observed by the Herschel satellite. The presence of an atomic and molecular in-flow in NGC~4418  would not be surprising. The galaxy and its interacting companion are in fact linked by an atomic hydrogen bridge \citep{klockner13}, and it is likely that part of that intergalactic material is being funneled into the potential well of the most massive galaxy of the pair.

The HI absorption may be tracing part of this atomic gas falling on the galaxy's nucleus along our line of sight.  From our observations it is not possible to detail the morphology of the in-flow, thus we cannot  confirm the material to be channeled toward the nucleus by a spiral arm rather than a spherically symmetric flow. However, because of the 0.$\!''$1  shift between SC and RC components, and the lack of significant blue-shifted emission, our CO 2--1 model does suggest the possible in-flow not to be spherically symmetric. Indeed, near-IR images of NGC~4418 by \cite{evans03} show dark radial features, which may be interpreted as dust lanes connecting the nucleus with the outer regions of the galaxy \citep[see Fig. 1,2 in ][]{evans03}. These structures may contain enough molecular and atomic gas to produce the observed red-shifted component. 
\subsubsection{Atomic and molecular mass}
\label{sec:mass}

The column density of the in-flowing atomic gas can be calculated from the observed HI opacity, following \cite{tools} :
\begin{equation}
\frac{N(\mathrm{H})}{\mathrm{cm^{-2}}} \simeq 1.82\times10^{18} \times \frac{T_\mathrm{s}}{\mathrm{K}} \times \int {\tau(v) \times \frac{\mathrm{d}v}{\mathrm{km~s^{-1}}}}, 
\end{equation} 
where $T_{\rm s}$ is the spin temperature of HI. If we assume a Gaussian line profile and an elliptical Gaussian absorbing region with major axis {\it maj} and minor axis {\it min}, the mass of HI is then
\begin{equation}\label{eq:mh1}
%\frac{M(H)}{M_\odot}=m_H \times \int{N(H) d\Omega} 
\frac{M(\mathrm{H})}{\mathrm{M}_\odot} \simeq 2.3 \times 10^{-2} \times \frac{T_{\rm s}}{\mathrm{K}} \times \tau_\mathrm{p} \times \frac{\Delta \mathrm{v}}{\mathrm{km~s^{-1}}} \frac{maj \times min}{\mathrm{pc^2}} ,
\end{equation} 
where $\tau_p$ is the peak opacity and $\Delta v$ is the velocity dispersion of the HI profile. 

The peak opacities of the two velocity components of HI are 0.5$\pm$0.08 (RC) and 0.2$\pm$0.1 (SC). If we assume the atomic hydrogen gas to have the same kinetic temperature of the CO~2--1 peak emission, i.e.  $T_{\rm s}\simeq$$T_\mathrm{CO}$=80 K, we find the HI column density towards the core of NGC~4418 to be 1.5$\pm$0.2$\times$10$^{22}$ and 5.8$\pm$2.9$\times$10$^{21}$ cm$^{-2}$,  for the RC and SC components respectively. This calculation assumes the atomic and molecular gas to be co-spatial and to have the same excitation temperature, which in general may not be true. In the case of a multi-phase clumpy medium, we would expect HI to have a higher temperature than CO, thus the values derived for the hydrogen column density should be considered as lower limits.

The size of the HI absorption is determined by the structure of the background continuum emission and it is only a lower limit to the extent of the atomic gas. In order to calculate the total HI mass, and to compare it to the total molecular mass, we assume the atomic gas to have the same spatial extent of CO~2--1 emission. In this case,  Eq. \ref{eq:mh1} gives an atomic hydrogen mass of 1.2$\pm$0.2$\times$10$^6$ and 4.9$\pm$2.9$\times$10$^5$ M$_\odot$, respectively for the RC and SC component.

In principle, the molecular hydrogen mass contained in the system can be calculated by applying an appropriate CO-H$_2$ conversion factor ($X_{\rm CO}$). Surveys of CO~1--0 emission in LIRGs \citep[e.g., ][]{solomon97, yao03} find values of $X_{\rm CO}$ which are three to ten times lower than what is found for Galactic giant molecular clouds (GMCs). This is usually attributed to the high turbulence in the cores of LIRGs. However, a recent study by \citet{papa12} suggests that in the warm and dense nuclei of LIRGs and ULIRGs, low-J CO lines may not be good tracers of the total molecular mass. For objects where observations of high-density tracers (e.g., high-J CO, HCN) are available, these authors find $X_{\rm CO}$ to be comparable to the Galactic value, and suggest that this could be a common property of LIRGs and ULIRGs nuclei. In the following analysis we will consider a conversion factor ranging from the classical ULIRG value of 1 M$_\odot$ (km s$^{-1}$ pc$^{-2}$)$^{-1}$ to a Galactic value of 4 M$_\odot$ (km s$^{-1}$ pc$^{-2}$)$^{-1}$ \citep{papa12}.

Single-dish observations by \citet{nascent} reveal a CO~2--1/1--0 brightness temperature ratio of roughly 0.5, when corrected for a source size of 0$''$.7. If we assume a conversion factor $X_{\rm CO}$=1-4 M$_\odot$ (km s$^{-1}$ pc$^{-2}$)$^{-1}$, by integrating on the CO~ 2--1 data-cube we find a molecular hydrogen mass of 0.6-2.5$\times$10$^8$ M$_\odot$ for the SC component and 1-4.2$\times$10$^7$ M$_\odot$ for RC. In the inner 100 pc, this corresponds to averaged H$_2$ column densities of 1.2-4.9$\times$10$^{23}$ cm$^{-2}$ for the red-shifted emission and  0.7-2.9$\times$10$^{24}$ cm$^{-2}$ for the systemic component. Given the uncertainties, these values agree well with the molecular mass of $\sim$10$^8$~M$_\odot$ estimated by \citet{sakamoto2013} from 850 $\mu$m continuum observations and HCN~4--3 kinematics in the inner 20 pc of the galaxy.

The molecular to atomic mass ratio $M$(H$_{\rm 2}$)/$M$(H) is 40-370 for SC and  12-300 for RC. Despite the large scatter due to the uncertainties on $X_{\rm CO}$, we see that $M$(H$_{\rm 2}$) is at least one order of magnitude greater than $M$(H) in both components, with a hint of a higher molecular fraction in the SC.  A similar result, less dependent on the assumptions used to derive molecular and atomic gas masses, can be derived from the spectra in Fig. \ref{fig:spec}. In general, the mass contained in the beam can be expressed as $M$(H)=$G$(H)$\times\Delta\mathrm{v}\times\alpha$ and $M$(H$_\mathrm{2}$)=$G$(CO)$\times\Delta\mathrm{v}\times\beta$, where G and $\Delta\mathrm{v}$ are the Gaussian amplitudes and line widths listed in Table \ref{tab:lines}, and $\alpha$ and $\beta$ are conversion factors (see e.g., Eq \ref{eq:mh1}). If we assume the excitation conditions to be the same for the two velocity components, $\alpha$ and $\beta$ are the same for SC and RC, and we obtain:

\begin{equation}
R_\mathrm{Mol}\equiv\frac{M(\mathrm{H_2})_\mathrm{SC}/M(\mathrm{H})_\mathrm{SC}}{M(\mathrm{H_2})_\mathrm{RC}/M(\mathrm{H})_\mathrm{RC}}=\frac{G(\mathrm{CO})_\mathrm{SC}\Delta\mathrm{v}}{G(\mathrm{H})_\mathrm{SC}\Delta\mathrm{v}}\times\frac{G(\mathrm{H})_\mathrm{RC}\Delta\mathrm{v}}{G(\mathrm{CO})_\mathrm{RC}\Delta\mathrm{v}},
\end{equation}

which is independent from conversion factors. By substituting values from Table \ref{tab:lines} we find $R_\mathrm{Mol}$=8$\pm$2, which confirms an higher molecular to atomic ratio in the SC component. If we assume the SC component to be dominated by the gas in the nucleus, and the RC to be mostly coming from the in-flowing envelope, these molecular ratios may suggest the presence of a density and pressure gradient, increasing toward the inner regions of the galaxy. However, because of the many uncertainties involved, and because we do not have the resolution to resolve the geometry of the in-flow, this is for now just an educated speculation based on the limited observations.

If we assume the HI absorption and CO 2--1 emission to be distributed on a similar scale of $\sim$80~pc, the total gas mass contained in the in-flow is $M_{\rm In}$=$M$(H$_2$)+$M$(H)=1-4.3$\times$10$^{7}$ M$_\odot$. Assuming a spherical geometry, the estimated in-flow mass would result in a gas density of $\sim$10$^3$~cm$^{-3}$, which is reasonable for CO 2--1 emitting gas (CO 2--1 has critical density n$_{\rm crit}\sim$10$^3$~cm$^{-3}$).

The mass flux depends on the geometry of the in-flowing gas. If we assume a projected size of the in-flow along the line of sight of $\Delta\ell\approx$80~pc (i.e., a spherical geometry), the total gas mass flux towards the nuclear regions of NGG~4418 can be calculated as $\dot{M}\approx M_{\rm In}\times\Delta\mathrm{v}\times\Delta\ell$, where $\Delta\mathrm{v}\approx$90 km s$^{-1}$ is the velocity shift between the SC and RC components. We find $\dot{M}\simeq$11-49~M$_\odot$~yr$^{-1}$. This value depends strongly on the assumed projected size of the in-flowing gas and must be considered as an order of magnitude estimate. A comparison with the in-flow model derived by \citet{galfonso2012} is presented in Section \ref{sec:her}.

\subsubsection{An outflow/in-flow scenario}

In a companion paper \citep{sakamoto2013}, a U-shaped red feature was found in an optical color index map, obtained as the ratio of (i$'$+z$'$)/(g$'$+r$'$) SDSS images. This is suggested to be associated to a minor-axis outflow, extending up to 10$''$ (1.5 kpc on the sky) on the northwestern side of the galaxy. The complex CO 2--1 and 3--2 velocity fields would then be the result of a mixed in-flow/outflow system. We speculate that the observed red-shifted CO and HI could be tracing an in-flow on the plane of the galaxy, along the line of sight, while the U-shaped absorption feature would be tracing an outflow perpendicular to the plane. For further discussion about the possible outflow, please refer to Section 3.6 in \citet{sakamoto2013}.

In the same paper (Section 3.4), an alternative explanation for the skewed CO~3--2 profile as the result of red-shifted self-absorption is also discussed. With the data at hand it is difficult to disentangle the two suggested models for CO emission, and we suggest the two interpretations as two equally probable scenarios. Both interpretations are consistent with the molecular in-flow detected by \citet{galfonso2012} with Herschel observations (see discussion in Section \ref{sec:her}).

\subsection{Molecular excitation}

\subsubsection{LTE analysis of the identified molecular lines}
\label{sec:lte}
\begin{figure*}
   \centering
   \includegraphics[width=.95\textwidth,keepaspectratio]{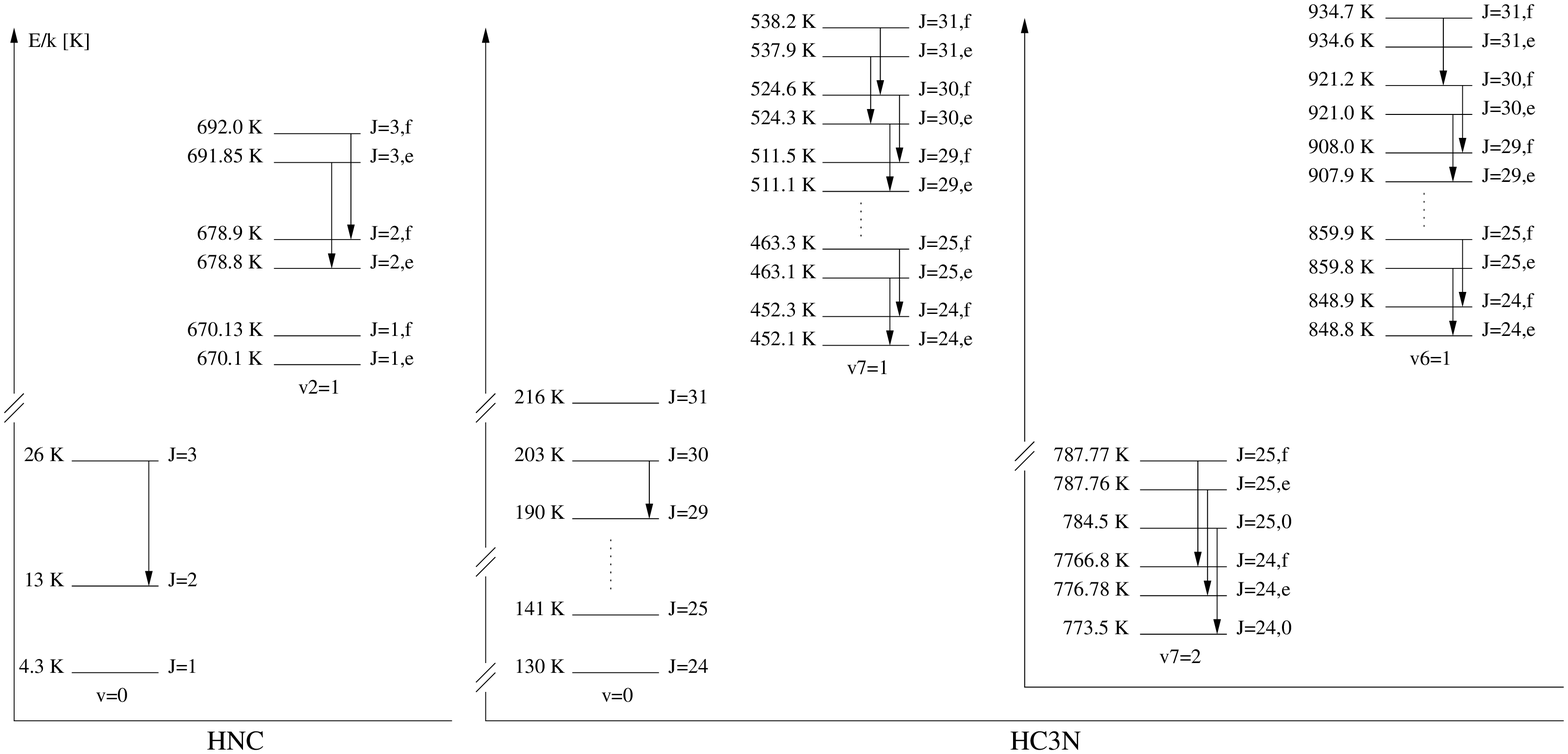}
      \caption{Rotational and vibrational levels of HNC and HC$_3$N. The solid arrows indicate the rotational transitions detected with the SMA.}
         \label{fig:ladder}
   \end{figure*}

The large velocity dispersions found for extragalactic molecular emission ($\sim$1000 km\,s$^{-1}$) often make the identification of spectral lines a challenging task. In the most extreme cases, when blending and line confusion make it almost impossible to reach an unambiguous Gaussian fit to the single spectral features, an effective technique consists in fitting a synthetic spectrum to the data \citep[e.g., ][]{martin2011}. Even in our case, despite the small (for an extragalactic object) line widths found in NGC~4418 ($\sim$100 km\,s$^{-1}$), significant line blending is present, especially for the faint transitions of vibrationally excited molecules. The observed spectra were thus fitted with a synthetic LTE model of the emission. 

Our model calculates the integrated brightness spectrum of a molecule, starting from an estimate of the molecule's column density $N$ and excitation temperature $T_\mathrm{ex}$, and assuming local thermal equilibrium. The transition parameters necessary for the calculation of LTE intensities were taken from the molecular line database {\it Splatalogue}.  

For the molecules which have many different transitions detected in the observed bands, a rotational-diagram analysis \citep[e.g, ][]{popdiag} of the blend-free lines is first attempted, to assign an appropriate starting value for $T_\mathrm{ex}$. For the molecules which have only one transition observed in the SMA bands, it is not possible to solve the degeneracy between column density and excitation temperature, and a fixed value is then chosen for  $T_\mathrm{ex}$. The integrated intensities calculated by the LTE code are then converted into brightness temperature spectral profiles by convolution with a Gaussian. The FWHM of the convolving line profile is set by Gaussian fitting the blend-free lines, or it is fixed at 100 km\,s$^{-1}$ when this is not possible. 

To compare the LTE model with our observations, we convert the flux density spectrum in Jy, extracted from the central synthesized beam, to main beam temperature units using

\begin{equation}\label{eq:tmb}
T_\mathrm{mb}/S=\frac{\lambda^2}{2k}\Omega_\mathrm{mb}^{-1}, 
\end{equation}

where $T_\mathrm{mb}$ is the main beam synthesized temperature, $S$ is the main beam flux density in Jy, and $\Omega_\mathrm{mb}$ is the synthesized beam  solid angle. The resulting K/Jy conversion factors for all the four observed SMA bands  are shown in Table~\ref{tab:cont}, column 5.  

The brightness temperature of the emission is related to the main beam temperature by

\begin{equation}\label{eq:tb}
T_\mathrm{b}=T_\mathrm{mb}\frac{\Omega_\mathrm{S\star mb}}{\Omega_\mathrm{S}},  
\end{equation}
where $\Omega_\mathrm{S\star mb}$ is the solid angle of the convolution between the source and the synthesized beam. For spatially resolved emission (CO, C$^{34}$S) we derive brightness temperature profiles, which can be directly compared to our synthetic spectrum. In the case of unresolved emission, we calculate lower limits to $T_\mathrm{b}$  by assuming a source size of one synthesized beam. Main beam temperatures and brightness temperatures, derived from equations \ref{eq:tmb} and \ref{eq:tb} for all detected molecular transitions, are shown in  Table~\ref{tab:lines}.  

The fitting of the LTE model to the data is done through minimization of the rms of the residuals, after subtracting the synthetic spectra from the observations. The excitation temperature of the species for which we do not have more than one transition in our spectra is fixed at $T_\mathrm{CO}$=80 K, which is the peak brightness temperature of the CO~2--1 emission. If we assume the CO $J$=2-1 transition to be optically thick,  $T_\mathrm{CO}$ is a good estimate of its excitation temperature, which at LTE is equal to the kinetic temperature of the molecular gas.

For  v$_6$=1 and v$_7$=1 vibrationally excited HC$_3$N, which show several transitions in our SMA bands, it is possible to estimate the rotational temperature via population diagram analysis, resulting in $T_\mathrm{ex}$=310 K for v$_6$=1 and $T_\mathrm{ex}$=160 K for v$_7$=1. 

The best-fit column densities deriving from our LTE model are listed in Table~\ref{tab:lte}. The model spectrum is shown in Fig. \ref{fig:spec}.

%\subsection{Molecular abundances}

\subsubsection{Vibrational excitation and IR pumping}
\label{sec:vibpump}
\begin{figure}
   \centering
   \includegraphics[width=.45\textwidth,keepaspectratio]{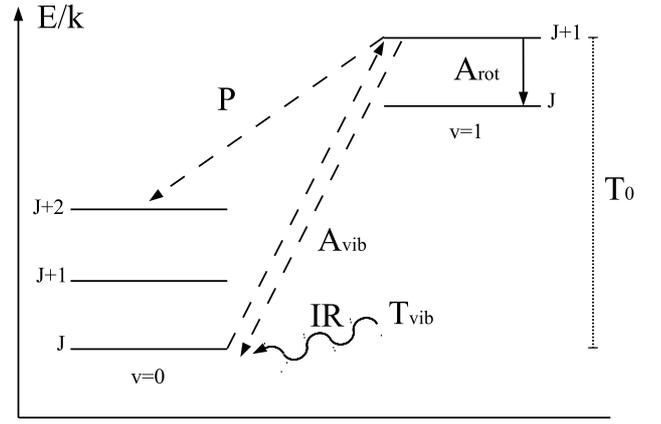}
      \caption{Pumping of rotational levels via IR vibrational excitation. Rotational and vibrational transitions are drawn respectively as solid and dashed arrows. See Section \ref{sec:vibpump} for discussion.}
         \label{fig:pump}
   \end{figure}
Vibrational excitation of molecules occurs via stretching and bending modes in the IR. The rotational spectrum of such vibrationally excited molecules can be observed at mm wavelengths and provides important clues about the physical conditions of the emitting region. Critical densities of vibrational transitions are usually higher than the maximum densities found in the molecular ISM \citep[$>$10$^8$ cm$^{-3}$, e.g., ][]{schilke92}, and their excitation is thus considered to be dominated by radiation. In the limit of optically thick vibrational transitions, the vibrational temperature derived from mm emission of vibrationally excited molecules is a good estimate of the temperature of the IR field in the emitting region.  For this reason, vibrationally excited molecular emission has been used by various authors as an extinction-free probe of the IR field in highly obscured objects, both in the Galaxy  \citep[e.g., ][]{wyr99, schilke03} and in extragalactic sources \citep[e.g., ][]{costagliola2010, sakamoto10}. 

Infrared excitation of vibrational modes may also play an important role in populating the rotational levels of the ground vibrational state \citep[e.g., ][]{carroll81}. Recent studies show that IR-pumping via vibrational modes may affect the intensity of rotational lines of even standard dense gas tracers such as HCN \citep{sakamoto10}, HNC \citep{schilke03} and HC$_3$N \citep{costagliola2010}, especially in regions with large IR fluxes.

Following \cite{carroll81}, the ($J$, v=0) state of a molecule can be excited by an IR photon to a ($J$+1, v=1) vibrationally excited state and then decay back either to the starting state or to ($J$+2,v=0). In the latter case, we say that the population of the $J$+2 level has been {\it pumped} by the IR field (See Fig. \ref{fig:pump}). The pumping rate can be estimated as 
\begin{equation}
\label{eq:pump}
P \simeq \frac{A_\mathrm{vib}}{e^{T_0/T_\mathrm{vib}}-1},
\end{equation}
where $A_\mathrm{vib}$ is the Einstein coefficient for the vibrational transition ($J$, v=0)$\leftrightarrow$($J$+1, v=1), and $T_{\rm 0}$ is the energy separation between v=1 and v=0 levels. This process will efficiently affect the excitation of the molecule if it can populate the $J$+2 level before ($J$+1, v=1) decays back to ($J$, v=1). In other words, following \cite{sakamoto10}, we must have
\begin{equation}
\label{eq:tp}
P > A_\mathrm{rot} \Leftrightarrow T_\mathrm{vib}>T_0/ln(1+A_\mathrm{vib}/A_\mathrm{rot}), 
\end{equation}
where $A_\mathrm{rot}$ is the Einstein coefficient for the rotational transition ($J$+1,v=1)$\rightarrow$($J$, v=1). The minimum IR brightness temperature needed for the pumping to be efficient is of the order of 100~K for the first levels of the most common dense gas tracers, as HCN, HCO$^+$ and CS \citep[e.g., ][]{sakamoto10}.

If we compare the pumping rate with the collisional excitation rate $C_{J-1,J}=n(H_2)q_{J-1,J}$, we can define a {\it pumping critical density}, $n_\mathrm{c}=P_{\ell u}/q_{J-1,J}$. For $n(H_2)<n_\mathrm{c}$ the radiative pumping process is more efficient than collisions at exciting the molecule. 

In Fig. \ref{fig:ladder} we show the rotational and vibrational energy ladders for HNC and HC$_3$N, and we mark the rotational transitions that we detected with the SMA. The detection of vibrationally excited HNC and HC$_3$N allows us to estimate their vibrational temperature and directly derive the conditions for IR pumping in NGC~4418.\\

\begin{figure}
 \centering
 \includegraphics[width=0.45\textwidth,keepaspectratio]{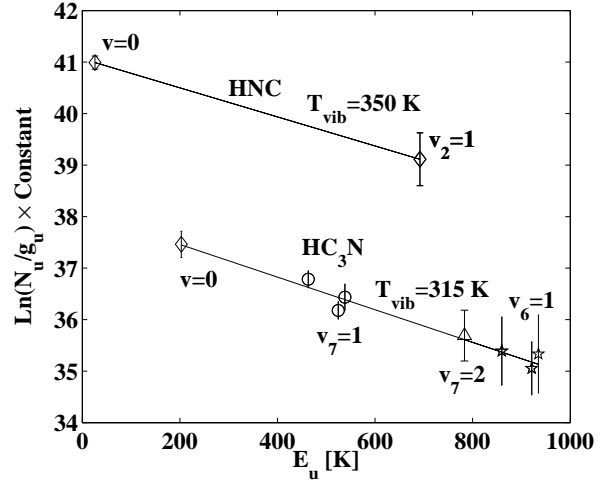}
 \caption{\label{fig:vib} Population diagram for vibrationally excited HNC and HC$_3$N. For HNC, the data points relative to the $J$=3, v=0, and v$_2$=1 states are shown as open $diamonds$. For HC$_3$N, the $J$=30, v=0 state is drawn as a $diamond$, while the v$_6$=1, v$_7$=1, and v$_7$=2 values, derived from all the observed transitions, are respectively shown as $stars$, $circles$ and a $triangle$. The solid lines show the vibrational temperature fit, while error bars represent 1-$\sigma$ uncertainties. For discussion, see Section \ref{sec:vibpump}.}
\end{figure}

{\it HNC: } The v$_2$=1 bending mode of HNC lies at 21 ${\rm \mu}$m,  about 700 K above the ground state,  and was first detected in space by \cite{schilke03} in the proto-planetary nebula CRL~618. Here we report the first (to our knowledge) extragalactic detection of mm emission from vibrationally excited HNC v$_2$=1. The intensity of the $J$=3-2, v$_\mathrm{2}$=1$e$ and v$_\mathrm{2}$=1$f$ lines, obtained from our LTE spectral fit (see Sections \ref{sec:hnc}, \ref{sec:lte}), was compared to the emission from the vibrational ground state via population-diagram analysis (see Fig. \ref{fig:vib}), which resulted in a vibrational temperature of 350 K. This value is larger than the vibrational temperature $\sim$230 K found by \cite{sakamoto10} for HCN 4-3 in NGC~4418. This may be due to an opacity effect, since in our analysis we are assuming optically thin transitions, which may not be true for the HNC, v=0 line. Also, because of a lower energy gap and a higher transition coefficient, the v$_2$=1 vibrational state of HNC is more easily excited than the corresponding state of HCN \citep[e.g., ][]{aalto07}. This could result in a larger volume of HNC to be vibrationally excited, which would show as an increase in vibrational temperature in the population diagram.

For the v$_2$=1 bending mode we have  $T_{\rm 0}$=669 K, and $A_\mathrm{vib}$=5.2 s$^{-1}$\citep{aalto07}. By assuming a rotational coefficient $A_{J=3\rightarrow2,\rm v_2=1}$=8$\times$10$^{-4}$ s$^{-1}$ (from {\it Splatalogue}), and applying  Eq. \ref{eq:tp}, we find that the minimum IR brightness temperature required to efficiently pump the HNC rotational levels via the v$_2$=1 vibration is 80 K. If we correctly interpret the estimated $T_\mathrm{vib}$ as a lower limit to the IR brightness temperature, it is thus possible that the molecule is being pumped. The pumping rate, estimated with Eq. \ref{eq:pump}, is P$\simeq$0.6 s$^{-1}$. The collisional coefficients for HNC transitions can be found in the {\it Leiden Atomic and Molecular Database}\footnote{\url{http://www.strw.leidenuniv.nl/~moldata/}} \citep{lambda}. If we consider a gas kinetic temperature of 80 K (see Section \ref{sec:lte}), we obtain a pumping critical density of $n_\mathrm{c}\simeq$10$^9$ cm $^{-3}$.  \\

{\it HC$_3$N: } A study of vibrationally excited HC$_3$N in NGC~4418 was first reported by \cite{costagliola2010}, who detected the v$_6$=1 and v$_7$=1 vibrationally excited states of the molecule through their rotational $J$=10-9, $J$=17-16, $J$=25-24 and $J$=28-27 mm-wave transitions. The two vibrational bending modes have IR transitions at 20 and 45 ${\rm \mu}$m, for the v$_6$=1 and v$_7$=1 respectively, with critical densities $>$10$^8$ cm$^{-3}$, and lie about 700 and 300 K above the vibrational ground level. 

From single-dish observations, \cite{costagliola2010} find that HC$_3$N is highly vibrationally excited, with vibrational temperatures as high as 500 K. Our population-diagram analysis confirms the high excitation, but with a lower vibrational temperature of 315 K (see Fig.  \ref{fig:vib}). These two results are consistent, given the large errors in the single-dish intensities deriving from calibration ($\sim$20$\%$) and baseline instabilities. The uncertainty of our vibrational temperature measurements was calculated with a Monte Carlo method, which resulted in an estimated 1-$\sigma$ error of 20$\%$. The intensity of the tentative detection of the v$_7$=2 vibrationally excited state of HC$_3$N is also well fitted by the vibrational temperature derived for the v$_6$=1 and v$_7$=1 states ($triangle$ in Fig. \ref{fig:vib}).

The vibrational coefficients are $A_\mathrm{vib}$=1.5$\times$10$^{-1}$ s$^{-1}$ for v$_6$=0--1, and  $A_\mathrm{vib}$=6$\times$10$^{-4}$ s$^{-1}$ for v$_7$=0--1 \citep{costagliola2010}. The rotational coefficients for the $J$=25--24, $J$=30--29, and $J$=31--30 are respectively $A_{J=25\rightarrow24, \mathrm{v}=1}$=9.3$\times$10$^{-4}$ s$^{-1}$, $A_{J=30\rightarrow29, \mathrm{v}=1}$=1.6$\times$10$^{-3}$ s$^{-1}$, and $A_{J=31\rightarrow30, \mathrm{v}=1}$=1.8$\times$10$^{-3}$ s$^{-1}$ (from {\it Splatalogue}). 

Using Eq. \ref{eq:tp}, we find that the minimum IR brightness temperature for vibrational pumping to be efficient ranges from 140 to 160 K for the v$_6$=1, and from 630 to 1100 K for the v$_7$=1 transitions. Considering the estimated $T_\mathrm{vib}\simeq$ 300 K, pumping via the v$_6$=1 may thus affect the molecules excitation, with a pumping rate P$\simeq$ 2$\times$10$^{-2}$ s$^{-1}$. If we assume a kinetic temperature of 80 K, the collisional coefficients for HC$_3$N in the 1 mm band are of the order of $q_{J-1,J}\simeq$ 10$^{-10}$ cm$^{3}$ s$^{-1}$, which results in a critical density for IR pumping of $n_\mathrm{c}\simeq$ 2$\times$10$^8$ cm$^{-3}$.\\

In the limit of optically thin rotational emission and efficient coupling between the IR field and the molecule's vibrational modes, $T_\mathrm{vib}$ is a good estimate of the IR field temperature in the region where the lines are formed \citep[e.g., ][]{costagliola2010, schilke03}. The fact that HNC and HC$_3$N show similar vibrational temperatures, may imply that rotational emission from  both vibrationally excited molecules is coming from a warm region, with IR temperatures of the order of 300 K. This is further discussed in Section \ref{sec:compact}.

We find that the pumping critical densities for both HNC and HC$_3$N have values $n_\mathrm{c}\geq$10$^8$ cm$^{-3}$, which are higher than the typical densities found in the dense interstellar medium (ISM). We can safely assume that only a small fraction of the gas will reside at  $n(H_2)>n_\mathrm{c}$, meaning that radiative pumping may be more efficient than collisions at populating the vibrational ground state of a detectable amount of  HNC and HC$_3$N in the nucleus of NGC~4418, or at least give a significant contribution to their excitation.

%\subsection{C$^{34}$S}

%-Does Kazushi have the CS 5-4?

\begin{table}
\caption{\label{tab:lte} Results from our LTE model of molecular emission in the SMA spectra. For HC$_3$N v$_6$=1 and v$_7$=1, the excitation temperature was calculated with the population diagram method. The excitation temperature was fixed at 80 K for the species for which only one transition was observed, while for HC$_3$N v$_7$=2 the rotational temperature was fixed at the value derived for HC$_3$N v$_7$=1 (160 K). The fixed parameters in the fit are marked with asterisks. Vibrational temperatures derived for HNC and HC$_3$N transitions are also shown. Uncertainties on the excitation temperatures are Monte Carlo estimates. See Section \ref{sec:lte} for discussion.  Because of the many uncertainties and assumptions involved, we advise extreme caution when interpreting temperatures and column densities derived by LTE models. Please see Section \ref{sec:limits} for discussion.}
% Results Continuum 
% Freq Flux(beam) Flux(tot) beamsize sourcesize  
\renewcommand{\arraystretch}{1.2}
\setlength{\tabcolsep}{12pt}
\begin{tabular}{lccc} 
\hline 
\hline 
& & &  \\ 
Molecule & T$_{\rm rot}$  & N  & T$_{\rm vib}$ \\
 & [K] & [cm$^{-2}$] &  [K] \\ 
 & &  &  \\ 
\hline 
& &  & \\ 
CO (SC) & 80* &  $>$ 1$\times$10$^{20}$ & \\
%\vspace{1mm}
CO (RC)	 & 80* &  5$\pm 1\times$10$^{18}$ & \\
%\vspace{1mm}
C$^{34}$S & 80* &  5$\pm 1\times$10$^{15}$ & \\
%\vspace{1mm}
HNC v=0	& 80* &  8$\pm 4\times$10$^{15}$ & \rdelim\}{2}{10mm}[ 350 $\pm$ 40]  \\
%\vspace{1mm}
HNC v$_2$=1 & 80* &  2$\pm 1\times$10$^{15}$ & \\
%\vspace{1mm}
HC$_3$N v=0  & 80* &  7$\pm 2\times$10$^{16}$ & \rdelim\}{4}{10mm}[ 315 $\pm$ 30] \\
%\vspace{1mm}
HC$_3$N v$_6$=1 & 310 $\pm$ 50 &  3.6$\pm 2\times$10$^{15}$ & \\
%\vspace{1mm}
HC$_3$N v$_7$=1 & 160 $\pm$ 30 &  6$\pm 3\times$10$^{15}$  & \\
%\vspace{1mm}
HC$_3$N v$_7$=2 & 160* &  4$\pm 2\times$10$^{15}$  & \\
%\vspace{1mm}
%CH$_3$OH & 80 &  $<${\bf ?}   & \\

& &  & \\ 
\hline
\hline
\end{tabular}

\end{table}

\subsubsection{Limitations of the LTE approach and main uncertainties}
\label{sec:limits}
In Table~\ref{tab:lte} we report the results of our excitation analysis, together with the uncertainties for the derived column densities and excitation temperatures. For the molecules which have only one detected transition (CO, CS, HNC (v=0), HC$_3$N (v=0)), the excitation temperature was fixed at 80 K, the peak of CO 2--1 emission (assuming CO to be optically thick). The uncertainty of 20$\%$ for the column density accounts only for the uncertainty in the data, and not for the uncertainty in the excitation temperature. The fundamental assumption here is that emission from CO, CS, HNC (v=0), and HC$_3$N (v=0), all come from the same ISM component at 80 K, which, although likely, may not be true. For these molecules, the uncertainties on the derived column densities should be thus interpreted as lower limits.

Our analysis assumes LTE, or at least that a single temperature can be used to describe the excitation of the molecular energy levels. In the case of rotational transitions, this implies collisions to be dominating the excitation, which is in general true for low-J transitions of most molecular tracers, these having critical densities $n_\mathrm{c}\leq$10$^5$--10$^6$ cm$^{-3}$. However, because of the steep increase of $n_\mathrm{c}$ with $J$, this does not hold for the higher transitions of dense gas tracers, which could become sub-thermally excited. This could affect the column densities and temperatures derived for vibrationally excited HC$_3$N.

However, in the presence of a strong IR field, it is likely that the excitation of the high-J levels will be dominated by radiation rather than collisions, and the derived excitation temperature will reflect the radiation temperature rather than the kinetic temperature of the gas. In this case, our LTE analysis would result in reasonably accurate estimates (50$\%$ uncertainty) of the column densities, and the derived excitation temperatures would be an estimate of the temperature of the IR source. This is further discussed in Section \ref{sec:tempstruct}.

In Section \ref{sec:vibpump} we find that IR pumping could play an important role in the excitation of the rotational levels of HNC and HC$_3$N. A non-LTE analysis, including vibrational excitation of HNC and HC$_3$N, would be needed to quantify this effect.

\subsection{Comparison with Herschel observations}
\label{sec:her}

A thorough study of molecular absorption in NGC~4418 with the Herschel PACS spectrometer was performed by \cite{galfonso2012}. These authors find that to explain the observed absorption lines from OH, HCN, H$_2$O and NH$_3$ a multi-component model is needed. Assuming a spherical geometry, the best fit is obtained with four different concentric components, labeled C$_\mathrm{hot}$, C$_\mathrm{core}$, C$_\mathrm{warm}$ and C$_\mathrm{ext}$.

The C$_\mathrm{hot}$ component is  contained in the innermost 5 pc of NGC~4418 nucleus and is needed to explain the mid-IR continuum emission. It has a high temperature of 350 K, but contains only a small fraction of the total molecular mass (8$\times$10$^3$M$_\odot$ for a total mass of roughly 10$^8$ M$_\odot$).  

The C$_\mathrm{core}$ and C$_\mathrm{warm}$ components extend from 5 to 30 pc, with temperatures of 150 K and 100 K, respectively. These components are needed to explain absorption from the high- and mid-lying levels of OH, HCN, H$_2$O and NH$_3$ and contain most of the molecular mass, roughly 10$^8$ M$_\odot$. 

The last more extended component, C$_\mathrm{ext}$, has a diameter of 200 pc,  temperatures ranging from 30 to 90 K and a mass of 1.8$\times$10$^7$ M$_\odot$. This extended envelope is responsible for the observed absorption from the low-lying levels of OH and from the [O I] transition at 63$\mu$m, which appear red-shifted from the systemic velocity of the galaxy by 100 km\,s$^{-1}$. The authors interpret this red-shifted absorbing envelope as a spherical in-flow, which would be driving molecular gas from the outer regions of NGC~4418 to its nucleus at a rate $\leq$ 12 M$_\odot$ yr$^{-1}$. 

The velocity and mass of our RC component, and the size of the CO~2--1 emission are consistent with the red-shifted absorbing envelope proposed by \cite{galfonso2012}. The greatest uncertainty is the geometry of the in-flow. In \cite{galfonso2012}, the in-flow is modeled as a collapsing spherical shell, while we model the SC component as a red-shifted spherical 3D Gaussian. The two different approaches may lead to differences in the estimated mass flux of the order of a few. The mass in-flow rate calculated from our CO~2--1 model is 11-49 M$_\odot$ yr$^{-1}$, which, given the uncertainties, is consistent with what was found by Herschel for the in-flow envelope. 

We notice that, while the shell model by \cite{galfonso2012} manages to explain most of the molecular absorption towards the core of NGC~4418, it does overestimate the blue-shifted component that one would expect in the case of a collapsing shell.  Our high-resolution CO~2--1 model reveals a spatially-resolved, red-shifted component, displaced by 0.$''$1 from the bulk of the molecular emission, and does not show any blue-shifted emission that would be compatible with a spherical in-flowing envelope. However the SC component is still barely resolved and we need yet higher-resolution observations to determine the exact geometry of the flow.

The layered temperature model suggested by  \cite{galfonso2012} is substantially confirmed by our interferometric observations, as discussed in the next Section.

\subsection{Temperature structure}
\label{sec:tempstruct}

 \begin{figure}
 \centering
 \includegraphics[width=0.4\textwidth,keepaspectratio]{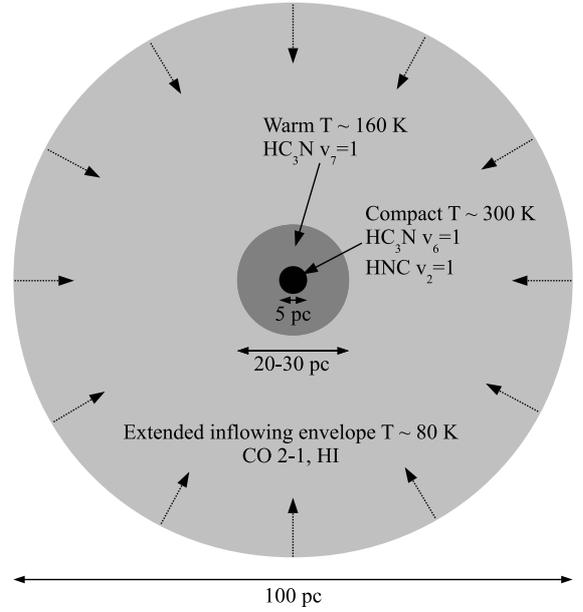}
 \caption{\label{fig:onion} Structure of the inner 100 pc of NGC~4418, inferred from our analysis of CO and HI dynamics, and from molecular excitation. The outer envelope, traced by CO and HI, has a temperature of 80 K (from CO brightness temperature) and shows clear in-flow signatures. A warmer component, at 160 K is traced by HC$_3$N v$_7$=1, and is consistent with the 20-30 pc-wide warm component derived by \cite{galfonso2012}. A compact, 300 K component of 5 pc in diameter, is traced by HC$_3$N v$_6$=1 and HNC  v$_2$=1. See also Table \ref{tab:lte}.}
\end{figure}

The LTE analysis of the detected molecular emission reveals a layered temperature structure, whih is most likely reflecting a steep temperature gradient in the nucleus of NGC~4418. We identify three main temperature components at 80 K, 160 K and 300 K (see Table~\ref{tab:lte}, and Fig. \ref{fig:onion}). 

The first component is derived from the CO~2--1 peak brightness temperature, and is assumed to be associated with the spatially resolved, extended molecular material. This value is comparable to the optically thick dust temperature of 85 K, found by \cite{evans03} by comparing IRAS fluxes at 100 and 60 $\mu$m, and to the temperature of the molecular gas (30-90 K) estimated by \cite{galfonso2012} for the outer envelope in their model of Herschel absorption. Also, the CO~2--1 peak is comparable to the CO~3--2 brightness temperature reported by \citet{sakamoto2013}. 

An intermediate temperature of 160 K derives from the rotational diagram analysis of HC$_3$N, v$_7$=1 emission. This value is consistent with what was found by  \cite{galfonso2012} for their C$_\mathrm{core}$ component (see Section \ref{sec:her}), and with the deconvolved brightness temperature of the 0.$''$1 continuum emission at 860 $\mu$m reported by \citet{sakamoto2013}.

A hot component at 300 K results both from the rotational excitation of the HC$_3$N, v$_6$=1 emission and from the vibrational excitation of HNC and HC$_3$N (see Fig. \ref{fig:vib} and Table~\ref{tab:lte}). Because of their high critical densities, the IR vibrational transitions of HNC and HC$_3$N can be excited only by radiation, and provide a direct, extinction-free probe of the IR field in the inner regions of NGC~4418. Assuming an efficient coupling between IR radiation and the molecular gas, the vibrational temperature found for the two molecular species, $T_\mathrm{vib}$=315-350 K,  is a good estimate of the temperature of the IR emission. 

The rotational temperature of HC$_3$N, v$_6$=1 ($\sim$310 K) is remarkably similar to the IR temperature inferred from our vibrational analysis. This may imply that the vibrationally excited molecular emission is emerging from a region where the IR field and the molecular gas are at equilibrium, i.e. an IR-thick source. In their model of Hershel molecular absorption, \cite{galfonso2012} need a hot, compact ($<$5 pc) component of $\sim$350 K in order to fit the observed mid-IR flux. This temperature is consistent with the hot component we detect with our excitation analysis. 

In principle, the hot component does not need to be a single compact source, but any distribution of 300~K gas with a total surface area $<$(5 pc)$^2$ contained in the beam would fit both SMA and Herschel observations. However, sub-mm observations by \citet{sakamoto2013} do show that most of the continuum dust emission is emerging from a compact region of less than 0.$''$1$\simeq$20~pc in diameter. We can thus speculate that both the mid-IR detected by Herschel and our vibrationally excited molecular emission are emerging from the same C$_\mathrm{hot}$, compact region in the core of NGC~4418.

\subsection{What is feeding at the center of NGC~4418?}
\label{sec:compact}

\subsubsection{A compact starburst ?}

If we assume that all the radio flux density at 1.4~GHz is due to synchrotron emission from supernovae, following \citet{condon92} we find a supernova rate of 0.04 yr$^\mathrm{-1}$ for the inner 100 pc of NGC~4418. Assuming that all stars with mass greater than 8 M$_\odot$ will evolve into a supernova,  the estimated supernova rate translates into a star formation rate SFR$\simeq$5 M$_\odot$ yr$^\mathrm{-1}$, for a Salpeter initial mass function \citep[IMF, ][]{salpeter55} between 0.1 and 120 M$_\odot$, and a continuous starburst older than 30 Myr. This value is consistent with an independent estimate of the star formation rate from the far-infrared (FIR) luminosity. Combining IRAS observations at 60 and 100 $\mu$, \cite{baan06} find an integrated (40-120 $\mu$m) FIR luminosity for NGC~4418 of Log($L_\mathrm{FIR}$/L$_\odot$)=10.63. If we apply the prescription from \cite{kennicutt98} to derive the SFR from the FIR luminosity, we find a galaxy-averaged star formation rate SFR$\simeq$7 M$_\odot$ yr$^\mathrm{-1}$, which is consistent with what was derived from the radio flux. {\it If the radio and IR luminosity is driven by star-formation, this galaxy is forming more stars than the whole Milky Way in a region of only 100 pc in diameter.} 

In Section \ref{sec:mass} we estimate for the inner 100~pc of NGC~4418 a molecular mass of 0.6-2.5$\times$10$^8$ M$_\odot$. If we assume the mass to be uniformly distributed on a region of 100~pc in diameter, we derive a gas surface density of $\mathrm{\Sigma_{gas}}$=0.8-3.2$\times$10$^{4}$ M$_\odot$ pc$^\mathrm{-2}$. Observations at higher spatial resolution by \citet{sakamoto2013} show that the bulk of the molecular gas is concentrated in a smaller region, of less than 20~pc in diameter. If we assume all the molecular gas to be contained in the inner 20~pc, the resulting surface density is 1.9-8$\times$10$^{5}$ M$_\odot$ pc$^\mathrm{-2}$. In the following discussion we will adopt a conservative value of $\mathrm{\Sigma_{gas}\sim}$10$^{4}$ M$_\odot$ pc$^\mathrm{-2}$.

Given such a gas surface density, the Kennicutt-Schmidt (KS) law \cite[e.g., ][]{kennicutt98} for normal galaxies and starbursts results in a star formation rate density of $\mathrm{\Sigma_{SFR}\simeq}$60 M$_\odot$ kpc$^\mathrm{-2}$ yr$^\mathrm{-1}$.  If we assume that star formation is uniformly distributed in the central 100 pc (the extent of the 1.4 GHz continuum), the SFR derived from radio and FIR fluxes imply an observed $\mathrm{\Sigma_{SFR}}$ which is roughly four times higher than what is derived from the KS law. Given the uncertainties, the two values can be considered as consistent, and the high star formation rate of NGC~4418 may be just a consequence of its extreme gas surface density.   

Both the radio and FIR methods discussed so far to derive the SFR assume a mature stellar population, resulting from a continuous star formation of at least 30-100 Myr. This assumption may not hold in NGC~4418, where the high luminosity-to-mass ratio, together with the high obscuration, suggest a much younger starburst activity. In \citet{sakamoto2013} SMA observations at 860~$\mu$m are used to derive a bolometric luminosity of 10$^{11}$ L$_\odot$ and a ratio $L_\mathrm{bol}/M_\mathrm{dyn}>$500~L$_\odot$~M$_\odot^{-1}$ for the inner 20~pc of the galaxy. These values, when compared to {\it Starburst99} simulations \citep{starburst99}, imply that if a starburst is powering the core of NGC~4418, its age must be less than 10~Myr and must have a SFR of more than 10~M$_\odot$~yr$^\mathrm{-1}$ (see Fig. 17 in \citet{sakamoto2013}).

At this rate, the total amount of molecular and neutral gas in the nucleus of NGC~4418 would be depleted in less than 30 Myr. However, we do have evidence of a molecular and atomic in-flow in NGC~4418. From our dynamical model, discussed in Section \ref{sec:co}, we estimate a mass flux towards the nucleus of the galaxy of 11-49~M$_\odot$~yr$^\mathrm{-1}$. Such a gas in-flow could be feeding the nuclear star formation, and it would considerably lengthen the life time of the starburst.

The high gas surface density derived for NGC~4418 resembles the extreme values ($\sim$5$\times$10$^4$ M$_\odot$ pc$^\mathrm{-2}$) found in much more massive ultra-luminous infrared galaxies \citep[ULIRGS, e.g, Arp~220, ][]{scoville97}. The inner 100 pc of NGC~4418 indeed share several properties with the western nucleus of the prototypical ULIRG,  Arp~220. If we consider the typical radio flux of supernova remnants (SNR) in Arp~220 \citep{batejat11}, the total radio emission at 1.4~GHz of NGC~4418 can be reproduced by a collection of 5 to 40 SNRs, depending on the age of the SNR. This would give a SNR surface density which is comparable with the one observed for the western nucleus of Arp~220. 

In a starburst scenario, the compact 300 K component derived from molecular vibrations could be emerging from the core of a super star cluster. In order to produce the observed IR luminosity, about 10$^6$ OB stars are needed \citep[e.g., ][]{hohle10}. These should be confined in a highly obscured region,  of only a few pc in size. The inferred IR luminosity and star density are higher than those commonly observed in obscured super-star clusters \citep[e.g., ][]{sauvage03}.

The maximum L/M  before the radiation pressure on dust starts blowing away the gas of a star-forming cloud is about 500-1000~L$_\odot$~M$_\odot^{-1}$ \citep[e.g., ][]{thompson05}. Our estimates of the hydrogen mass in the inner 100~pc of NGC~4418 result into a L/M=500-2000~L$_\odot$~M$_\odot^{-1}$, where the uncertainty mainly depends on the $X_{\rm CO}$ conversion factor (either Galactic or ULIRG, see Section \ref{sec:mass}). {\it Given such a large uncertainty,  we cannot exclude an extreme star-cluster, emitting at the maximum possible luminosity, as a possible power source at the center of NGC~4418.} We speculate that such an extreme super star cluster may be a result of the unusually high gas density in the core of NGC~4418. 

The upper end of our L/M estimate is not consistent with star formation to be powering the IR emission from the inner 100 pc. We investigate the possibility of a non-stellar contribution in the next Section.

\subsubsection{An active galactic nucleus ?}

Our excitation analysis confirms the layered thermal structure suggested by \cite{galfonso2012}, and provides a direct detection of the hot, compact source in NGC~4418 (see Section \ref{sec:tempstruct}). Following \cite{evans03}, if we assume all the IR continuum to be coming from a 300 K optically thick source, we obtain an upper limit for the size of the emitting region of just 5 pc. The existence of a compact AGN core in NGC~4418 has been proposed by several authors, mainly because of the high surface brightness in the IR \citep[e.g., ][]{spoon_01}, and the  lack of significant PAH emission in the mid-IR \citep{imanishi04}. The hot, compact core traced by vibrationally excited lines could be the first direct evidence of such a compact source.

Observations with the Chandra observatory by \cite{maiolino03} show that the hard X-ray flux from NGC~4418 is very faint, consistent with a very low power, Compton-thick AGN. However, because of the poor photon statistics, the authors claim that such an interpretation has to be regarded as tentative. From our CO 2--1 observations we estimate a molecular hydrogen column in the central SMA beam of 6$\times$10$^{24}$ cm$^{-2}$, which is consistent with the value of N(H)$>$10$^{25}$ cm$^{-2}$ derived by  \citet{sakamoto2013} for the 0.$''$1 dust continuum core. Such high column densities could cause an AGN to become Compton-thick. 

An empirical correlation between the bulge K-band luminosity and black hole mass has been found both for Seyfert and normal galaxies  \citep[e.g., ][]{peng06}. An estimate of the mass of a super-massive black hole in the center of NGC~4418 can be thus obtained from observations in the K band. The total K-band luminosity observed by the 2MASS mission \citep{2mass} is $L_{\rm K}\simeq$2$\times$10$^{10}$ L$_\odot$. This corresponds to a  black hole mass of roughly $M_{\rm BH}\simeq$7$\times$10$^6$ M$_\odot$ (see Fig. 4 in \cite{peng06}). As discussed in \citet{sakamoto2013}, the luminosity-to-mass ratio observed for the nucleus of NGC~4418 could be reproduced by an AGN of 1$\times$10$^7$ M$_\odot$, radiating at 30\% of its Eddington limit. 

For radio-quiet AGNs, \cite{nelson2000} found that the radio power and black hole mass are correlated. According to their Fig. 2, a black hole with the mass of 7$\times$10$^6$ M$_\odot$, should have a radio luminosity at 1.4~GHz of 3$\times$10$^{21}$ W Hz$^{-1}$, which is roughly ten times higher than the total radio luminosity observed for NGC~4418. For some reason, if an AGN is present in the center of the galaxy, it should be radio deficient. However, the empirical correlations just mentioned have quite a large scatter (especially the $L_{\rm K}/M_{\rm BH}$ relation) and their theoretical interpretation is still debated. Also, it is possible that the radio emission is being affected by synchrotron self-absorption or free–free absorption.

The low HCO$^+$/HCN $J$=1--0 line ratio observed in NGC~4418 has been interpreted by some authors as a signature of AGN activity \citep[e.g., ][]{kohno2001, krips08}. Standard models of X-ray-dominated chemistry \citep[e.g.,][]{meijerink07}, which should be dominant in an AGN environment, show instead an enhancement of HCO$^+$ emission due to the large ionization, and a consequently large HCO$^+$/HCN ratio. Also, the presence of bright HC$_3$N emission is difficult to explain in an AGN scenario, the molecule being easily destroyed by reactions with atomic and molecular ions and by UV radiation \citep[e.g., ][]{turner98}. In a high resolution study of the chemistry of IC~342, \citet{meier05} indeed find that HC$_3$N emission follows the 3~mm continuum dust emission, and anti-correlates with regions of intense UV radiation, such as photo-dissociation regions (PDRs).  However, recent models by \cite{harada10}, show that high-temperature reactions ($T>$400 K) can enhance HC$_3$N abundances on the plane of a dense AGN accretion torus. The bright emission from vibrationally excited  HC$_3$N may be thus coming from a small region in the accretion disk. Models show that the size of such a region should be of a few pc, which would be consistent with what we estimate from HC$_3$N and HCN vibrational temperatures.

\section{Conclusions}

We have observed the compact obscured nucleus of NGC~4418 with MERLIN and the SMA, at radio and mm wavelengths. The radio emission at 1.4 and 5~GHz is spatially resolved, with an estimated source size of 80 and 25 pc.  The CO 2--1 emission and HI absorption are fit by a two-component model, with the two Gaussian components, having centroid velocities of 2090 and 2180 km\,s$^{-1}$, and hydrogen masses of  0.6-2.5$\times$10$^8$ and 1.1-4.3$\times$10$^7$ M$_\odot$. {\it These observations confirm the existence of in-flowing gas towards the nucleus of NGC~4418, which was first detected in absorption by \citet{galfonso2012} with Herschel. We estimate the molecular and atomic in-flow to be feeding the central activity at a rate of 11-49~M$_\odot$ yr$^{-1}$.} 

Vibrationally excited HC$_3$N and HNC are detected, with vibrational temperatures of roughly 300 K. Molecular excitation in NGC~4418 is consistent with a layered temperature structure, with three main components at 80, 160 and 300 K. These may reflect a steep temperature gradient, rather than three distinct temperatures. For the hot component we estimate a source size of less than 5 pc. We discuss the radiative pumping of the HC$_3$N and HNC rotational levels via vibrational transitions, and we find that this may play an important role in the excitation of the molecules. 

If the radio and FIR emission is produced by star formation, the high L/M ratio implies a star formation rate of $\sim$10~M$_\odot$ yr$^{-1}$, and a starburst age of less than 10 Myr. Instead, the radio continuum may be emerging from an AGN. Using an empirical relation between the black hole mass and K-band luminosity, we estimate a $M_{\rm BH}\simeq$7$\times$10$^6$ M$_\odot$ for the putative black hole in the center of NGC~4418. The total 1.4~GHz flux density derived from MERLIN observations is roughly 10 times lower than what would be expected for an AGN of this mass. 

Despite  these high angular resolution observations, the nature of the central energy source in NGC~4418 is still unclear. However, this study provides some new constrains to the properties of the nuclear starburst or AGN: 
\begin{itemize}
\item If a compact starburst is producing the observed IR luminosity, it has to be at least as extreme as the one observed in Arp~220. If we consider the estimated gas depletion time of 30~Myr, and the high $L/M>$500~L$_\odot$~M$_\odot^{-1}$, the starburst must be from 3 to 10 Myr old. 
\item If an AGN is feeding at the center of NGC~4418, this must be Compton-thick, so that its presence is not clearly revealed by hard X-rays observations. The low radio flux observed, may be due to an intrinsic property of the AGN, or to highly compact free-free absorption.
\end{itemize}

Wide-field VLA observations suggest that NGC~4418 is capturing gas from its HI-rich companion \citep{klockner13}. This gas appears to be entering NGC~4418 from the north-eastern side of the galaxy. How the captured gas gets from the 16~kpc-scale atomic bridge into the inner 100 pc of NGC~4418 is unclear.

The compactness of NGC~4418 nucleus appears to be key to the observed extreme properties of gas and dust. Similar gas properties are observed in ULIRGs, such as Arp~220, but these are usually interacting systems of much larger mass. 

%{\bf If we want to solve the intricate puzzle of the compact nucleus of NGC~4418, observations at even higher spatial resolution are needed. We plan to use the new capabilities of the ALMA array to study the gas dynamics at sub-mm wavelengths, and in a paper now in preparation we will use an ALMA spectral scan of NGC~4418 to study in detail the molecular excitation. Radio VLBI observations would also be extremely valuable, because they would allow us to resolve the supernova remnants, and would give very strict constraints on the existence of an AGN in the center of NGC~4418. }

\begin{acknowledgements}

F. C. acknowledges support by the Spanish MINECO through grant AYA 2012-38491- C02-02, co-funded with FEDER funds.
      
\end{acknowledgements}

\bibliographystyle{aa} % style aa.bst
\bibliography{bibliototale}

\begin{thebibliography}{55}
\expandafter\ifx\csname natexlab\endcsname\relax\def\natexlab#1{#1}\fi

\bibitem[{{Aalto} {et~al.}(2007{\natexlab{a}}){Aalto}, {Monje}, \&
  {Mart{\'{\i}}n}}]{nascent}
{Aalto}, S., {Monje}, R., \& {Mart{\'{\i}}n}, S. 2007{\natexlab{a}}, \aap, 475,
  479

\bibitem[{{Aalto} {et~al.}(2007{\natexlab{b}}){Aalto}, {Spaans}, {Wiedner}, \&
  {H{\"u}ttemeister}}]{aalto07}
{Aalto}, S., {Spaans}, M., {Wiedner}, M.~C., \& {H{\"u}ttemeister}, S.
  2007{\natexlab{b}}, \aap, 464, 193

\bibitem[{{Baan} {et~al.}(2008){Baan}, {Henkel}, {Loenen}, {Baudry}, \&
  {Wiklind}}]{baan08}
{Baan}, W.~A., {Henkel}, C., {Loenen}, A.~F., {Baudry}, A., \& {Wiklind}, T.
  2008, \aap, 477, 747

\bibitem[{{Baan} \& {Kl{\"o}ckner}(2006)}]{baan06}
{Baan}, W.~A. \& {Kl{\"o}ckner}, H.-R. 2006, \aap, 449, 559

\bibitem[{{Batejat} {et~al.}(2011){Batejat}, {Conway}, {Hurley}, {Parra},
  {Diamond}, {Lonsdale}, \& {Lonsdale}}]{batejat11}
{Batejat}, F., {Conway}, J.~E., {Hurley}, R., {et~al.} 2011, \apj, 740, 95

\bibitem[{{Carroll} \& {Goldsmith}(1981)}]{carroll81}
{Carroll}, T.~J. \& {Goldsmith}, P.~F. 1981, \apj, 245, 891

\bibitem[{{Condon}(1992)}]{condon92}
{Condon}, J.~J. 1992, \araa, 30, 575

\bibitem[{{Costagliola} \& {Aalto}(2010)}]{costagliola2010}
{Costagliola}, F. \& {Aalto}, S. 2010, \aap, 515, A71

\bibitem[{{Costagliola} {et~al.}(2011){Costagliola}, {Aalto}, {Rodriguez},
  {Muller}, {Spoon}, {Mart{\'{\i}}n}, {Per{\'e}z-Torres}, {Alberdi},
  {Lindberg}, {Batejat}, {J{\"u}tte}, {van der Werf}, \&
  {Lahuis}}]{costagliola11}
{Costagliola}, F., {Aalto}, S., {Rodriguez}, M.~I., {et~al.} 2011, \aap, 528,
  A30

\bibitem[{{Elmegreen} \& {Efremov}(1997)}]{elmegreen97}
{Elmegreen}, B.~G. \& {Efremov}, Y.~N. 1997, \apj, 480, 235

\bibitem[{{Evans} {et~al.}(2003){Evans}, {Becklin}, {Scoville}, {Neugebauer},
  {Soifer}, {Matthews}, {Ressler}, {Werner}, \& {Rieke}}]{evans03}
{Evans}, A.~S., {Becklin}, E.~E., {Scoville}, N.~Z., {et~al.} 2003, \aj, 125,
  2341

\bibitem[{{Goldsmith} \& {Langer}(1999)}]{popdiag}
{Goldsmith}, P.~F. \& {Langer}, W.~D. 1999, \apj, 517, 209

\bibitem[{{Gonz{\'a}lez-Alfonso} {et~al.}(2012){Gonz{\'a}lez-Alfonso},
  {Fischer}, {Graci{\'a}-Carpio}, {Sturm}, {Hailey-Dunsheath}, {Lutz},
  {Poglitsch}, {Contursi}, {Feuchtgruber}, {Veilleux}, {Spoon}, {Verma},
  {Christopher}, {Davies}, {Sternberg}, {Genzel}, \& {Tacconi}}]{galfonso2012}
{Gonz{\'a}lez-Alfonso}, E., {Fischer}, J., {Graci{\'a}-Carpio}, J., {et~al.}
  2012, \aap, 541, A4

\bibitem[{{Harada} {et~al.}(2010){Harada}, {Herbst}, \& {Wakelam}}]{harada10}
{Harada}, N., {Herbst}, E., \& {Wakelam}, V. 2010, \apj, 721, 1570

\bibitem[{{Hohle} {et~al.}(2010){Hohle}, {Neuh{\"a}user}, \&
  {Schutz}}]{hohle10}
{Hohle}, M.~M., {Neuh{\"a}user}, R., \& {Schutz}, B.~F. 2010, Astronomische
  Nachrichten, 331, 349

\bibitem[{{Imanishi} {et~al.}(2004){Imanishi}, {Nakanishi}, {Kuno}, \&
  {Kohno}}]{imanishi04}
{Imanishi}, M., {Nakanishi}, K., {Kuno}, N., \& {Kohno}, K. 2004, \aj, 128,
  2037

\bibitem[{{Kennicutt}(1998)}]{kennicutt98}
{Kennicutt}, Jr., R.~C. 1998, \araa, 36, 189

\bibitem[{{Kl\"ockner} {et~al.}(2013){Kl\"ockner}, {Beswick}, {Aalto},
  {Costagliola}, {Muller}, {Sakamoto}, \& {Mart{\'{\i}}n}}]{klockner13}
{Kl\"ockner}, H.~R., {Beswick}, R., {Aalto}, S., {et~al.} 2013, in preparation

\bibitem[{{Kohno} {et~al.}(2001){Kohno}, {Matsushita}, {Vila-Vilar{\'o}},
  {Okumura}, {Shibatsuka}, {Okiura}, {Ishizuki}, \& {Kawabe}}]{kohno2001}
{Kohno}, K., {Matsushita}, S., {Vila-Vilar{\'o}}, B., {et~al.} 2001, in
  Astronomical Society of the Pacific Conference Series, Vol. 249, The Central
  Kiloparsec of Starbursts and AGN: The La Palma Connection, ed. {J.~H.~Knapen,
  J.~E.~Beckman, I.~Shlosman, \& T.~J.~Mahoney}, 672

\bibitem[{{Krips} {et~al.}(2008){Krips}, {Neri}, {Garc{\'{\i}}a-Burillo},
  {Mart{\'{\i}}n}, {Combes}, {Graci{\'a}-Carpio}, \& {Eckart}}]{krips08}
{Krips}, M., {Neri}, R., {Garc{\'{\i}}a-Burillo}, S., {et~al.} 2008, \apj, 677,
  262

\bibitem[{{Leitherer} {et~al.}(1999){Leitherer}, {Schaerer}, {Goldader},
  {Gonz{\'a}lez Delgado}, {Robert}, {Kune}, {de Mello}, {Devost}, \&
  {Heckman}}]{starburst99}
{Leitherer}, C., {Schaerer}, D., {Goldader}, J.~D., {et~al.} 1999, \apjs, 123,
  3

\bibitem[{{Maiolino} {et~al.}(2003){Maiolino}, {Comastri}, {Gilli}, {Nagar},
  {Bianchi}, {B{\"o}ker}, {Colbert}, {Krabbe}, {Marconi}, {Matt}, \&
  {Salvati}}]{maiolino03}
{Maiolino}, R., {Comastri}, A., {Gilli}, R., {et~al.} 2003, \mnras, 344, L59

\bibitem[{{Mart{\'{\i}}n} {et~al.}(2011){Mart{\'{\i}}n}, {Krips},
  {Mart{\'{\i}}n-Pintado}, {Aalto}, {Zhao}, {Peck}, {Petitpas}, {Monje},
  {Greve}, \& {An}}]{martin2011}
{Mart{\'{\i}}n}, S., {Krips}, M., {Mart{\'{\i}}n-Pintado}, J., {et~al.} 2011,
  \aap, 527, A36

\bibitem[{{Meier} \& {Turner}(2005)}]{meier05}
{Meier}, D.~S. \& {Turner}, J.~L. 2005, \apj, 618, 259

\bibitem[{Meijerink {et~al.}({2007})Meijerink, Spaans, \& Israel}]{meijerink07}
Meijerink, R., Spaans, M., \& Israel, F.~P. {2007}, {A\&A}, {461}, 793

\bibitem[{{M{\"u}ller} {et~al.}(2005){M{\"u}ller}, {Schl{\"o}der}, {Stutzki},
  \& {Winnewisser}}]{cdms}
{M{\"u}ller}, H.~S.~P., {Schl{\"o}der}, F., {Stutzki}, J., \& {Winnewisser}, G.
  2005, Journal of Molecular Structure, 742, 215

\bibitem[{{Nelson}(2000)}]{nelson2000}
{Nelson}, C.~H. 2000, \apjl, 544, L91

\bibitem[{{Papadopoulos} {et~al.}(2012){Papadopoulos}, {van der Werf},
  {Xilouris}, {Isaak}, \& {Gao}}]{papa12}
{Papadopoulos}, P.~P., {van der Werf}, P., {Xilouris}, E., {Isaak}, K.~G., \&
  {Gao}, Y. 2012, \apj, 751, 10

\bibitem[{{Peng} {et~al.}(2006){Peng}, {Gu}, {Melnick}, \& {Zhao}}]{peng06}
{Peng}, Z., {Gu}, Q., {Melnick}, J., \& {Zhao}, Y. 2006, \aap, 453, 863

\bibitem[{{P{\'e}rez-Torres} {et~al.}(2009){P{\'e}rez-Torres},
  {Romero-Ca{\~n}izales}, {Alberdi}, \& {Polatidis}}]{miguel09}
{P{\'e}rez-Torres}, M.~A., {Romero-Ca{\~n}izales}, C., {Alberdi}, A., \&
  {Polatidis}, A. 2009, \aap, 507, L17

\bibitem[{{Pickett} {et~al.}(1998){Pickett}, {Poynter}, {Cohen}, {Delitsky},
  {Pearson}, \& {M{\"u}ller}}]{jpl}
{Pickett}, H.~M., {Poynter}, R.~L., {Cohen}, E.~A., {et~al.} 1998, \jqsrt, 60,
  883

\bibitem[{{Roche} {et~al.}(1986){Roche}, {Aitken}, {Smith}, \&
  {James}}]{roche86}
{Roche}, P.~F., {Aitken}, D.~K., {Smith}, C.~H., \& {James}, S.~D. 1986,
  \mnras, 218, 19P

\bibitem[{{Rohlfs} \& {Wilson}(1996)}]{tools}
{Rohlfs}, K. \& {Wilson}, T.~L. 1996, {Tools of Radio Astronomy}, ed. {Rohlfs,
  K.~\& Wilson, T.~L.}

\bibitem[{{Roussel} {et~al.}(2003){Roussel}, {Helou}, {Beck}, {Condon},
  {Bosma}, {Matthews}, \& {Jarrett}}]{roussel03}
{Roussel}, H., {Helou}, G., {Beck}, R., {et~al.} 2003, \apj, 593, 733

\bibitem[{{Sakamoto} {et~al.}(2013){Sakamoto}, {Aalto}, {Costagliola},
  {Mart{\'{\i}}n}, {Ohyama}, {Wiedner}, \& {Wilner}}]{sakamoto2013}
{Sakamoto}, K., {Aalto}, S., {Costagliola}, F., {et~al.} 2013, \apj, 764, 42

\bibitem[{{Sakamoto} {et~al.}(2010){Sakamoto}, {Aalto}, {Evans}, {Wiedner}, \&
  {Wilner}}]{sakamoto10}
{Sakamoto}, K., {Aalto}, S., {Evans}, A.~S., {Wiedner}, M.~C., \& {Wilner},
  D.~J. 2010, \apjl, 725, L228

\bibitem[{{Salpeter}(1955)}]{salpeter55}
{Salpeter}, E.~E. 1955, \apj, 121, 161

\bibitem[{{Sanders} {et~al.}(2003){Sanders}, {Mazzarella}, {Kim}, {Surace}, \&
  {Soifer}}]{sanders03}
{Sanders}, D.~B., {Mazzarella}, J.~M., {Kim}, D.-C., {Surace}, J.~A., \&
  {Soifer}, B.~T. 2003, \aj, 126, 1607

\bibitem[{Sanders \& Mirabel({1996})}]{sanders_96}
Sanders, D.~B. \& Mirabel, I.~F. {1996}, {Annu. Rev. Astron. Astrophys.}, {34},
  749

\bibitem[{{Sauvage} \& {Plante}(2003)}]{sauvage03}
{Sauvage}, M. \& {Plante}, S. 2003, \apss, 284, 941

\bibitem[{{Schilke} {et~al.}(2003){Schilke}, {Comito}, \&
  {Thorwirth}}]{schilke03}
{Schilke}, P., {Comito}, C., \& {Thorwirth}, S. 2003, \apjl, 582, L101

\bibitem[{{Schilke} {et~al.}(1992){Schilke}, {Walmsley}, {Pineau Des Forets},
  {Roueff}, {Flower}, \& {Guilloteau}}]{schilke92}
{Schilke}, P., {Walmsley}, C.~M., {Pineau Des Forets}, G., {et~al.} 1992, \aap,
  256, 595

\bibitem[{{Sch{\"o}ier} {et~al.}(2005){Sch{\"o}ier}, {van der Tak}, {van
  Dishoeck}, \& {Black}}]{lambda}
{Sch{\"o}ier}, F.~L., {van der Tak}, F.~F.~S., {van Dishoeck}, E.~F., \&
  {Black}, J.~H. 2005, \aap, 432, 369

\bibitem[{{Scoville} {et~al.}(1997){Scoville}, {Yun}, \& {Bryant}}]{scoville97}
{Scoville}, N.~Z., {Yun}, M.~S., \& {Bryant}, P.~M. 1997, \apj, 484, 702

\bibitem[{{Skrutskie} {et~al.}(2006){Skrutskie}, {Cutri}, {Stiening},
  {Weinberg}, {Schneider}, {Carpenter}, {Beichman}, {Capps}, {Chester},
  {Elias}, {Huchra}, {Liebert}, {Lonsdale}, {Monet}, {Price}, {Seitzer},
  {Jarrett}, {Kirkpatrick}, {Gizis}, {Howard}, {Evans}, {Fowler}, {Fullmer},
  {Hurt}, {Light}, {Kopan}, {Marsh}, {McCallon}, {Tam}, {Van Dyk}, \&
  {Wheelock}}]{2mass}
{Skrutskie}, M.~F., {Cutri}, R.~M., {Stiening}, R., {et~al.} 2006, \aj, 131,
  1163

\bibitem[{{Solomon} {et~al.}(1997){Solomon}, {Downes}, {Radford}, \&
  {Barrett}}]{solomon97}
{Solomon}, P.~M., {Downes}, D., {Radford}, S.~J.~E., \& {Barrett}, J.~W. 1997,
  \apj, 478, 144

\bibitem[{Spaans \& Meijerink({2004})}]{spaans_05}
Spaans, M. \& Meijerink, R. {2004}, {Astrophysics and Space Science}, {295},
  239

\bibitem[{{Spoon} {et~al.}(2001){Spoon}, {Keane}, {Tielens}, {Lutz}, \&
  {Moorwood}}]{spoon_01}
{Spoon}, H.~W.~W., {Keane}, J.~V., {Tielens}, A.~G.~G.~M., {Lutz}, D., \&
  {Moorwood}, A.~F.~M. 2001, \aap, 365, L353

\bibitem[{{Spoon} {et~al.}(2007){Spoon}, {Marshall}, {Houck}, {Elitzur}, {Hao},
  {Armus}, {Brandl}, \& {Charmandaris}}]{spoon07}
{Spoon}, H.~W.~W., {Marshall}, J.~A., {Houck}, J.~R., {et~al.} 2007, \apjl,
  654, L49

\bibitem[{{Swinbank} {et~al.}(2011){Swinbank}, {Papadopoulos}, {Cox}, {Krips},
  {Ivison}, {Smail}, {Thomson}, {Neri}, {Richard}, \& {Ebeling}}]{swinbank11}
{Swinbank}, A.~M., {Papadopoulos}, P.~P., {Cox}, P., {et~al.} 2011, \apj, 742,
  11

\bibitem[{{Thompson} {et~al.}(2005){Thompson}, {Quataert}, \&
  {Murray}}]{thompson05}
{Thompson}, T.~A., {Quataert}, E., \& {Murray}, N. 2005, \apj, 630, 167

\bibitem[{{Turner} {et~al.}(1998){Turner}, {Lee}, \& {Herbst}}]{turner98}
{Turner}, B.~E., {Lee}, H., \& {Herbst}, E. 1998, \apjs, 115, 91

\bibitem[{{Wyrowski} {et~al.}(1999){Wyrowski}, {Schilke}, \&
  {Walmsley}}]{wyr99}
{Wyrowski}, F., {Schilke}, P., \& {Walmsley}, C.~M. 1999, \aap, 341, 882

\bibitem[{{Yamada} \& {Creswell}(1986)}]{yamada86}
{Yamada}, K.~M.~T. \& {Creswell}, R.~A. 1986, Journal of Molecular
  Spectroscopy, 116, 384

\bibitem[{{Yao} {et~al.}(2003){Yao}, {Seaquist}, {Kuno}, \& {Dunne}}]{yao03}
{Yao}, L., {Seaquist}, E.~R., {Kuno}, N., \& {Dunne}, L. 2003, \apj, 588, 771

\end{thebibliography}

\begin{appendix}

%\section{MERLIN dirty beams}

 %\begin{figure*}
 %  \centering
 %  \includegraphics[angle=-90,width=0.4\textwidth,keepaspectratio]{figures/HI-beam.ps}
%	\includegraphics[angle=-90,width=0.4\textwidth,keepaspectratio]{figures/cont5GHz-beam.ps}
 %     \caption{MERLIN dirty beams.}
 %        \label{fig:mdirt}
 %  \end{figure*}

\section{Visibility fits}

\begin{figure*}
   \centering
   \includegraphics[width=.9\textwidth,keepaspectratio]{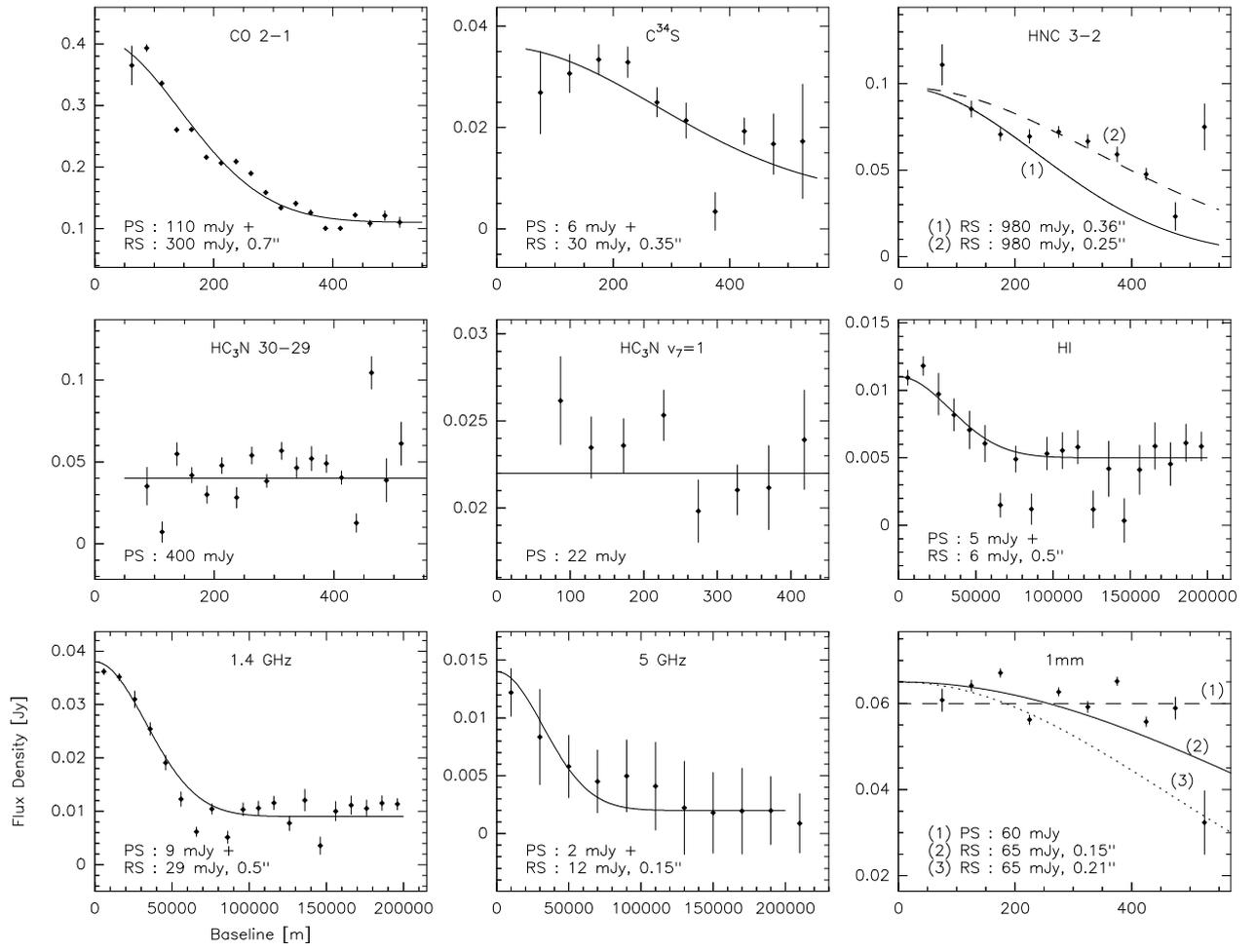}
      \caption{Circle-averaged visibility fits. The best fit parameters are labeled as $PS$ for the point source component, and $RS$ for the resolved component. The FWHM of the resolved component is given in arc-seconds. For HNC and the continuum at 1~mm, the averaged visibilities are compared with models of different source size, which are labeled as $(1)...(3)$.}
         \label{fig:sizes}
   \end{figure*}

To derive source sizes for the observed emission, a simple model of a point source and a circular Gaussian was fit to the circle-averaged visibilities. These were derived by an azimuthal average around the u-v plane center, with the {\sc uv\_circle} routine of the {\sc mapping} software. Averaged visibilities are plotted in Fig. \ref{fig:sizes}, on the x-axis we show the distance from the u-v center $\sqrt{u^2+v^2}$ in meters. The fitted parameters for the point source and Gaussian components are labeled as $PS$ and $RS$, respectively. 

\end{appendix}

\end{document}